\documentclass[journal]{IEEEtran}
\usepackage{amsmath,amsfonts}
\usepackage{algorithmic}
\usepackage{algorithm}
\usepackage{array}
\usepackage{textcomp}
\usepackage{stfloats}
\usepackage{url}
\usepackage{verbatim}
\usepackage{cite}
\usepackage{hyperref}
\usepackage{multirow}
\usepackage{longtable}
\usepackage{ltxtable}
\usepackage{booktabs}
\usepackage{authblk}
\usepackage{hhline}
\usepackage{amssymb}
\usepackage{lineno}
\usepackage[utf8]{inputenc}
\usepackage{placeins}
\usepackage{units}
\usepackage{setspace}
\usepackage{enumitem}
\usepackage{cases}
\usepackage[dvipsnames]{xcolor}
\usepackage{lscape} 
\usepackage{pdflscape}
\usepackage{ltxtable}
\usepackage{graphicx}
\usepackage{subcaption}
\usepackage{caption}
\usepackage{multicol}
\usepackage{afterpage}
\usepackage{nicematrix}
\usepackage{amsmath}

\renewcommand{\thetable}{\arabic{table}}

\begin{document}

\title{Urban Flood Mapping Using Satellite Synthetic Aperture Radar Data: A Review of Characteristics, Approaches and Datasets}

\author{Jie Zhao~\IEEEmembership{Member,~IEEE},
        Ming Li, 
        Yu Li, 
        Patrick Matgen, 
        Marco Chini~\IEEEmembership{Senior Member,~IEEE}

\thanks{This work is supported the Luxembourg National Research Fund (FNR) through the Swift project (Grant no. INTER/ANR/23/17800438) and the project "Provision of an Automated, Global, Satellite-based Flood Monitoring Product for the Copernicus Emergency Management Service" (GFM), Contract No. 939866-IPR-2020 for the European Commission’s Joint Research Centre (EC-JRC). (Corresponding author: Jie Zhao, Ming Li).}
\thanks{Jie Zhao works within the Department of Aerospace and Geodesy, Data Science in Earth Observation, Technical University of Munich, Arcisstraße 21, 80333 Munich, Germany, and was within the Remote Sensing Research Group, Department of Geodesy and Geoinformation, Vienna University of Technology, 1040 Vienna, Austria.}
\thanks{Ming Li works within Institute of Geodesy and Photogrammetry, ETH Zurich, 8039 Zurich, Switzerland, Wuhan University, LIESMARS and 4DiXplorer AG, Zurich, Switzerland}
\thanks{Yu Li, Patrick Matgen and Marco Chini work within the Environmental Research and Innovation Department, Luxembourg Institute of Science and Technology (LIST), Luxembourg. (e-mail: jie.zhao@tum.de, mingli39@ethz.ch/lisouming@whu.edu.cn, yu.li@list.lu, patrick.matgen@list.lu, marco.chini@list.lu)}}

\markboth{Journal of \LaTeX\ Class Files,~Vol.~14, No.~8, August~2021}%
{Shell \MakeLowercase{\textit{et al.}}: A Sample Article Using IEEEtran.cls for IEEE Journals}


\maketitle

\begin{abstract}

Understanding the extent of urban flooding is crucial for assessing building damage, casualties and economic losses. Synthetic Aperture Radar (SAR) technology offers significant advantages for mapping flooded urban areas due to its ability to collect data regardless weather and solar illumination conditions. However, the wide range of existing methods makes it difficult to choose the best approach for a specific situation and to identify future research directions. Therefore, this study provides a comprehensive review of current research on urban flood mapping using SAR data, summarizing key characteristics of floodwater in SAR images and outlining various approaches from scientific articles. Additionally, we provide a brief overview of the advantages and disadvantages of each method category, along with guidance on selecting the most suitable approach for different scenarios. This study focuses on the challenges and advancements in SAR-based urban flood mapping. It specifically addresses the limitations of spatial and temporal resolution in SAR data and discusses the essential pre-processing steps. Moreover, the article explores the potential benefits of Polarimetric SAR (PolSAR) techniques and uncertainty analysis for future research. Furthermore, it highlights a lack of open-access SAR datasets for urban flood mapping, hindering development in advanced deep learning-based methods. Besides, we evaluated the Technology Readiness Levels (TRLs) of urban flood mapping techniques to identify challenges and future research areas. Finally, the study explores the practical applications of SAR-based urban flood mapping in both the private and public sectors and provides a comprehensive overview of the benefits and potential impact of these methods.

\end{abstract}


\section{Introduction}
\label{sec:intro}

\IEEEPARstart{F}{loods} are among the most destructive natural disasters worldwide, whose frequency and magnitude are increasing due to growing heavy precipitation resulting from global climate change~\cite{Caretta2022water}. Generally, floods can occur in different land cover/land use areas, such as open areas, agricultural areas/vegetated areas and urban areas. Among those areas, urban flooding is a growing concern due to rapid urbanization and increasing flood risk. More than half of the world's population now lives in cities~\cite{Dodman2022Climate}, and many of these urban areas are expanding into flood-prone zones. More specifically, human settlements, ranging from villages to megacities, expanded globally by 84.5\% between 1985 and 2015, and settlement moved into high hazard flood zones outpacing growth in flood-safe areas~\cite{rentschler2023global}. Thus, urban floods can lead to high death tolls, enormous financial losses and hinder socially sustainable development. For example,~\cite{tellman2021satellite} analyzed over ten thousand cloud-free optical images from 2000 to 2018 and found that a growing number of people are at risk of flooding. This trend is alarming, with economic losses estimated at \$651 billion USD during this period. The devatating floods in Kenya and Libya in 2023 further underscore the urgent need for action. In Kenya, 46 people lost their lives, and 58,000 were displaced~\cite{reliefweb2023}. The situation in Libya was even more tragic, with 5,300 deaths and thousand missing~\cite{floodlistlibya2023}. Additionally, the UK's Suffolk region faced flooding in October 2023 due to Storm Babet, affecting 1,260 properties~\cite{floodlist2023}. Urban floods can also have severe consequences beyond property damage and loss of life. The COVID-19 pandemic highlighted the increased risk of disease transmission during and after floods~\cite{han2021urban}. Clearly, developing effective methods for urban flood mapping is essential for supporting flood risk managers, emergency responders, and local governments in their efforts to monitor, forecast, and manage flood events. 

Synthetic Aperture Radar (SAR) is a powerful tool for the extraction of flooded urban from space and it has several advantages in comparison with optical data. Indeed, SAR systems use longer wavelengths of the electromagnetic spectrum, allowing cloud penetration and weather independent image acquisition. Furthermore, SAR systems are active sensors, transmitting and receiving their own electromagnetic impulses, which allows operations independent of daylight. Currently, thanks to the growing number of SAR satellite constellations (e.g., Sentinel-1, COSMO-SkyMed, TerraSAR-X, ALOS-2, RCM, SAOCOM and GF3), a massive amount of diverse SAR data is now available to map water bodies at global scale. Moreover, the rapid development of commercial SAR satellites constellations (e.g., Capella, HiSea, ICEYE, Synspective,Umbra-X) provides a possibility for monitoring of ground features and their changes over time at increasing temporal resolution, making near-real time flood mapping envisageable in the near future. 

Although SAR data offers significant advantages for detecting urban floods, interpreting the radar signal can be complex due to various factors. Densely built-up areas often have challenges like mutual shadowing between buildings, multiple reflections, building orientation, and reflections from flat roofs. These factors, along with scattering from other elements like windows, can make it difficult to accurately map floods in urban areas~\cite{pierdicca2018flood}. For example, SAR backscatter intensity values vary depending on the difference between floodwater depth and the surrounding buildings’ heights~\cite{pulvirenti2011flood, mason2014detection}: SAR backscatter can significantly increase during urban floods due to the stronger double-bounce interaction between the specular reflecting water surface and the buildings' façades if buildings are partially submerged; When buildings are fully submerged by floodwater, the radar backscatter from the area can decrease significantly due to the smooth water surface reflecting the signal directly back to the satellite~\cite{zhao2022urban}. The radar backscatter from urban floods can vary in different ways. To address this, interferometric SAR (InSAR) technology, which analyzes both the signal strength and phase, has been shown to be a valuable tool for flood mapping~\cite{chini2012analysis,chini2019sentinel,li2019urban,pulvirenti2017detection,pulvirenti2020insar,pelich2022mapping,zhao2022urban}. This is bacause the presence of floodwater in urban areas often causes a rapid decrease in InSAR coherence. It should be noted that the InSAR coherence is calculated using a moving window of a fixed size, leading to comparatively lower spatial resolution when compared to SAR intensity. Additionally, changes in coherence in flooded urban areas often occur around buildings due to the way radar signals bounce off structures. In some cases, the coherence might not change significantly due to flooding, especially in areas with large, flat buildings and narrow streets. This could lead to missed detections of urban floods, as noted in previous studies~\cite{zhao2022urban}. Therefore, \cite{ohki2020flood,pulvirenti2020insar} argue that it is important to consider the phase information in addition to coherence in dense urban areas because phase is more sensitive to subtle changes caused by floodwater. 

It is important to note that the SAR characteristics discussed so far primarily focus on buildings surrounded by water. This is different from the broader concept of urban flooding, which can also include flooded streets, open spaces, and other areas within cities. The term 'urban' is complex and can have different meanings depending on the context. While it generally refers to densely populated areas with infrastructure, the specific definition can vary widely. For example, it could refer to a small village, a town, or a magacity~\cite{urban2015blog,urban2024Nationalgeographic}. In principle, urban regions should contain all of the surroundings, including all of the built-ups, vegetation, parks, airports, small lakes and roadways. In the remote sensing field, we have several different terms/products for the definition of urban areas, such as the impervious surfaces/areas in global land cover map (FROMFROM-GLC10)~\cite{gong2013finer,gong2019stable} including roads, driveways, sidewalks, parking lots, rooftops, etc~\cite{weng2012remote}, global urban footprint (GUF) / world settlement footprint (WSF) covering all built-ups from megacities to small villages \cite{esch2017breaking,esch2018we,marconcini2021understanding}, the global human settlement layer (GHSL) proposed by the European Joint Research Center (JRC) containing all the built-ups classification from Sentinel-2 images~\cite{schiavina2022ghsl}, and the urban/built-up layers in several global land cover/land use maps \cite{tsendbazar2021towards}. To the best of our knowledge, the definitions of urban floods are not always the same among different studies: some methods are described that urban floods are able to be delineated with relatively low accuracies in flooded urban areas, as their methods are developed mainly for flooded open areas/bare soils~\cite{schlaffer2015flood,liu2018water}. These approaches can only identify flooded public areas in urban settings that are not surrounded by buildings; some urban flood mapping approaches are proposed focusing on the built-up regions~\cite{zhao2022urban,li2019urban} and some others investigated floodwater mapping in residential regions ~\cite{chaabani2018flood}. To be clear, the general concept of urban flooding usually encompasses two scenarios: buildings surrounded by floodwater, and flooding that only affects main streets and public areas without encircling any buildings. In the second scenario, these areas can be considered similar to rural flooding, as there are no significant obstacles like buildings. Methods for detecting flooded open areas are summarized in several studies such as \cite{schumann2023flood,dasgupta2018flood}. However, due to the side-looking nature of the SAR sensor and complex environments, detecting flooded buildings remains a challenging task (details explained in Section\ref{sec:Characteristics}). Therefore, our study concentrates on urban flood mapping studies focusing on \textit{\textbf{flooded built-ups}} where the buildings are surrounded by floodwater based on the characteristics in SAR data. 

In recent years, significant research has focused on developing algorithms to extract urban flood information from SAR data. This is driven by advancements in computer vision, SAR technology, and the growing need for real-time flood monitoring. Several studies have explored the potential of SAR systems for detecting urban floods. These reports summarize the knowledge about the relationships between the sensors’ parameters (i.e., wavelength, polarization, incidence angle, interferometry) and environmental conditions (e.g. dielectric constant, topographic effects, rain/snow, etc.) \cite{amitrano2024flood,schumann2023flood, dasgupta2018flood,manavalan2018review,pierdicca2018flood,schumann2015microwave}. However, no specific review has taken place on SAR-based urban flood mapping. 

This article aims to provide a comprehensive overview of SAR-based urban flood mapping, addressing the following key areas:

\begin{itemize}

\item  Sensor Characteristics: Understanding how sensor parameters (i.e., wavelength, polarization, incidence angle, interferometry) and land cover types affect the SAR signal for urban flood detection.
\item  Classification Algorithms: Evaluating the latest methods for extracting urban flood information from SAR data, including their advantages and limitations.
\item  Future trends: Exploring emerging technologies and challenges that will shape the future of SAR-based urban flood mapping.
\item  Case Studies: Examining sepecific examples of urban flood events capture by Sentinel-1 to illustrate the application of SAR-based methods.
\item  Practical Implications: Highlighting the importance of urban flood mapping for both private and public sectors.

\end{itemize}

The article is composed of seven distinct sections. Section \ref{sec:Summary_method} provides an overview of the literature that serves as the foundation for all the analyses discussed in this work. Section \ref{sec:Characteristics} presents a concise summary of the properties of SAR for urban floods. It outlines the connection between SAR features such as wavelength, polarisation, and incidence angle, and the environmental factors like water level and urban form that are relevant to urban floods. Section \ref{sec:method_section} categorises current approaches into four distinct categories, each possessing its own advantages and limitations. Section \ref{sec:Challenges} examines the previously described findings and demonstrates upcoming trends, along with the necessity to enhance urban flood mapping. Additionally, this section allows users to access a comprehensive list of urban flood events, along with a detailed description of the dataset utilised. Section \ref{sec:Application} delineates the practical uses of SAR-based techniques for mapping urban floods, encompassing flood risk management and the endeavour to achieve sustainable development goals. Section \ref{sec:conclusions} presents conclusions.

\section{Review method on SAR-based urban flood mapping studies}
\label{sec:Summary_method}

\begin{figure}[!tbp]
    \centering
    \includegraphics[width=\linewidth]{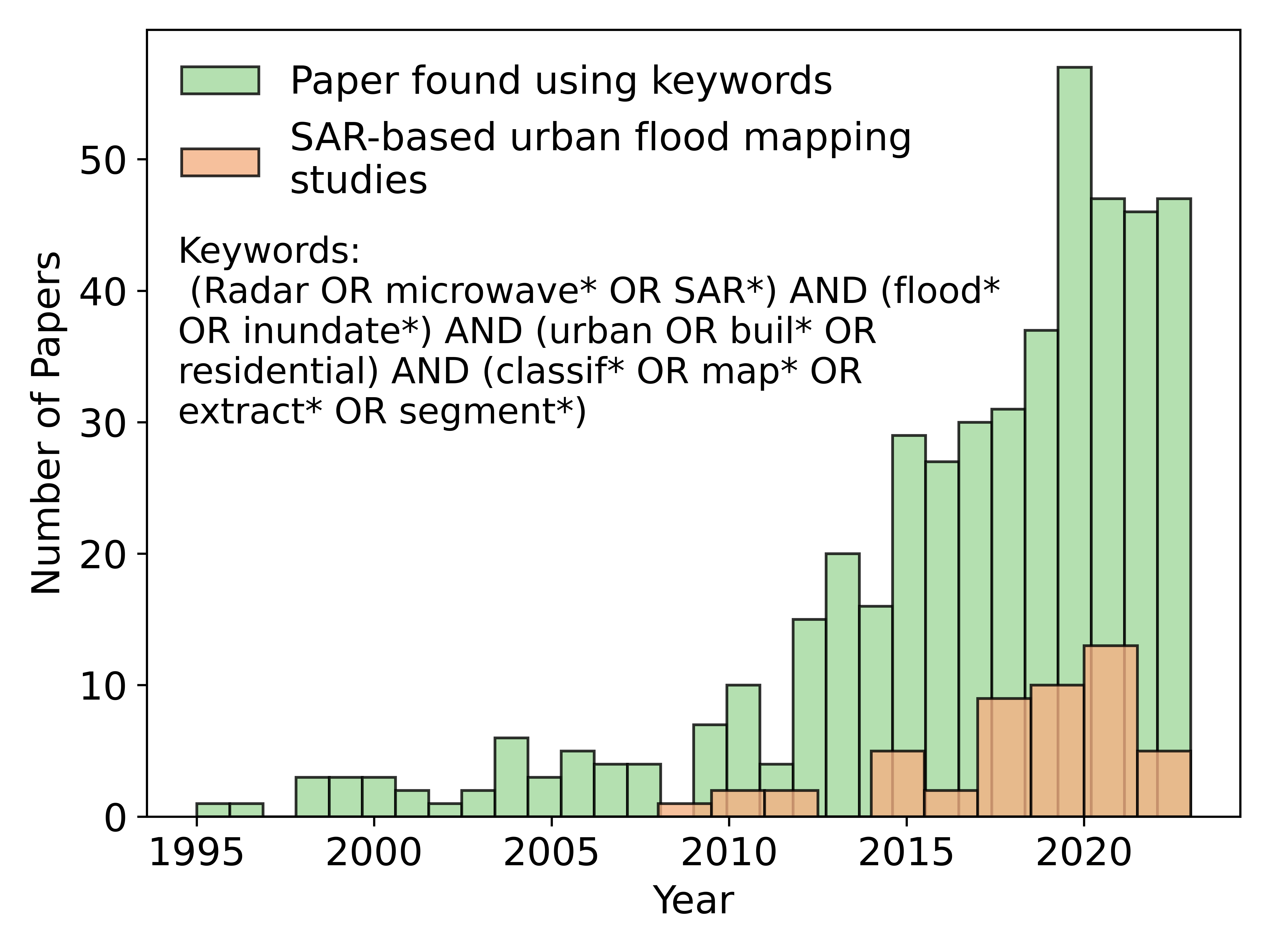}
    \caption{Information of studies selection for this review: only 10.6\% of the papers have been selected,while 89.4\% of the studies were excluded.}
    \label{fig:Review_info}
\end{figure}

In order to identify the relevant publications in the field, a combination of keywords linked to urban flood mapping was applied as structured queries for the period from 1996 to 2023 using \href{https://apps.webofscience.com/}{Web of Science} as a search engine, where only articles, proceeding papers and review articles are considered. Initially, 461 results have been selected. Further selection was conducted based on attentive analysis of the content of these articles: studies that do not refer to SAR data application in flooded built-ups were excluded. Then, the resulting set of studies containing 49 articles was used as a basic for the study (see Figure \ref{fig:Review_info}). These studies were then analyzed based on various factors, including the type of sensor used, polarization, spatial resolution, flood type, data modality, and the Technology Readiness Level (TRL) of each method, as summarized in Table~\ref{tab:list_paper}.

\begin{onecolumn}

\begin{landscape}
\begin{longtable}{c|c|c|c|c|c|c}

  \caption{Information of selected SAR-based urban flood mapping studies. (FB represents flooded urban areas, FO represents flooded open areas, and FV represents flooded vegetation. TRL: Technology Readiness Level. DSM: Digital Surface Model. DEM: Digital Elevation Model.) \centering} 
  
  \label{tab:list_paper}
  \\
  \toprule[2pt] \hline

  \textbf{Reference} & \textbf{Sensor}&  \textbf{Polarization} & \textbf{Resolution [m]} & \textbf{Flood type} & \textbf{Modality}  &\textbf{TRL} \\ 
  \hline  \endhead
  
  \cite{pierdicca2008integrating} & ERS-1    & VV & 12.5 & FB, FO        & SAR intensity,Land Cover/DEM     &5 \\ \hline
  \cite{pulvirenti2010analysis}   & COSMO-SkyMed      & HH & 5    & FB, FO, FV       &SAR intensity     & 5\\ \hline

  \cite{mason2010flood}            & TerraSAR-X & HH & 3    & FB         &SAR intensity, LiDAR DSM        &5\\ \hline

  \cite{chini2012analysis}         & COSMO-SkyMed & HH & 3    & FB, FO           &SAR intensity, InSAR coherence, land cover map       & 2 \\ \hline

  \cite{mason2012near}            & TerraSAR-X&  HH & 3    & FB          &SAR intensity, LiDAR DSM      & 5 \\ \hline

  \cite{pierdicca2014flood}        & COSMO-SkyMed &  HH & 3    & FB         &SAR intensity, InSAR coherence      &5 \\\hline

 \multirow{2}{*}{\cite{tanguy2014development}}      & \multirow{2}{*}{RADARSAT-2}&  \multirow{2}{*}{HH,HV} & \multirow{2}{*}{8}    & \multirow{2}{*}{FB}    &SAR intensity, Hydraulic model      & \multirow{2}{*}{5}\\

       &  &   &      &      & derived probability of flooding map        &  \\\hline

 \cite{mason2014detection}            & TerraSAR-X&  HH & 3    & FB          &SAR intensity, LiDAR DSM       &5 \\\hline
 \cite{pulvirenti2015integration}& COSMO-SkyMed&  HH & 3    & FB          &SAR intensity, InSAR coherence, DEM /land cover map      &5\\ \hline

 \cite{iervolino2015flooding}            & TerraSAR-X & HH & 3    & FB          &SAR intensity, Gauges information      &4 \\ \hline

 \cite{chini2016sar}            & Sentinel-1& VV, VH & 20    & FB, FO, FV      &SAR intensity, InSAR coherence    & 5 \\ \hline
 \cite{pulvirenti2016use}        &COSMO-SkyMed& HH & 3    & FB, FV       &SAR intensity, InSAR coherence     &5 \\ \hline

 \cite{tanguy2017river}        &RADARSAT-2& HH & 8    & FB       &SAR intensity, Flood return period      & 5 \\ \hline
 \cite{pulvirenti2017detection}  &Sentinel-1 & - & 15    & FB       &SAR intensity, InSAR coherence, land cover map   & 5\\ \hline

 \cite{kwak2017new}  &ALOS-2/PALSAR-2&  HH & 3    & FB      &SAR intensity      & 5  \\ \hline

 \cite{liu2018detection}  &ALOS-2/PALSAR-2& HH & 4.2 $\times$  2.9    & FB      &SAR intensity, Land cover map       &5 \\ \hline

 \cite{chaabani2018flood}  &TerraSAR-X & HH & 3    & FB, FO      &SAR intensity, InSAR coherence   &5 \\ \hline

 \cite{kwak2018effect}  &ALOS-2/PALSAR-2 & HH & 2.5    & FB       &SAR intensity, building footprint    & 5 \\ \hline

 \multirow{2}{*}{\cite{mason2018robust}}  &TerraSAR-X,  & \multirow{2}{*}{HH} & 3    & \multirow{2}{*}{FB}      & \multirow{2}{*}{SAR intensity, LiDAR DSM}   & \multirow{2}{*}{5} \\
 & COSMO-SkyMed        &  & 2.5  &             &     &    \\ \hline

 \cite{chini2018monitoring}  &Sentinel-1 & VV & 20   & FB, FO      &SAR intensity, InSAR coherence         & 5 \\ \hline

 \cite{natsuaki2018band}  &ALOS-2/PALSAR-2 &HH & 3   & FB, FO     &SAR intensity, InSAR coherence      &5 \\ \hline

 \cite{vanama2019change}  &Sentinel-1&  VV & -   & FB, FO      &SAR intensity, STRM water body mask      &5 \\ \hline

 \cite{pulvirenti2019flood}  &Sentinel-1 & - & 15   & FB      &InSAR coherence       &5 \\ \hline

 \cite{lin2019urban}  &Sentinel-1&  VV & 15   & FB, FO, FV       &SAR intensity      & 5 \\ \hline

 \cite{li2019urban}  &TerraSAR-X&  HH & 1.2 $\times$ 3.3  & FB, FO, FV       &SAR intensity, InSAR coherence      & 5 \\ \hline
 
 \cite{chini2019probabilistic}  &Sentinel-1 & VV & 20   & FB, FO     &SAR intensity, InSAR coherence      &5 \\ \hline
 
 \cite{chini2019sentinel}  &Sentinel-1& VV, VH & 5/20   & FB, FO      &SAR intensity, InSAR coherence    & 5 \\ \hline
 
 \multirow{2}{*}{\cite{ohki2019flood}}  &\multirow{2}{*}{ALOS-2/PALSAR-2 } 	       & VV, HH, & \multirow{2}{*}{3/6}   & \multirow{2}{*}{FB, FO}       &\multirow{2}{*}{SAR intensity, InSAR coherence, Building Polygon}        & \multirow{2}{*}{5} \\
 &      &VH, HV    &  &           &    &   \\ \hline
 
 \cite{ohki2019floodIGARSS}  &ALOS-2/PALSAR-2 & HH,VH & 5  & FB       &InSAR phase      &5 \\ \hline

 \multirow{2}{*}{\cite{li2019urbanRS}}  &Sentinel-1,&  VV, & 20,  & \multirow{2}{*}{FB, FO}       & \multirow{2}{*}{SAR intensity, InSAR coherence}        \multirow{2}{*}{5} \\ 
    &ALOS-2/PALSAR-2         & HH & 2.5 &            &       &  \\ \hline

\cite{benoudjit2019novel}  &TerraSAR-X  & HH & 3  & FB, FO        &SAR intensity, optical        &5\\  \hline

\multirow{2}{*}{\cite{olthof2020testing}}  &RADARSAT-2, & HH, HV & 8/12.5  & \multirow{2}{*}{FB, FO}     &SAR intensity, & \multirow{2}{*}{5} \\
 &Sentinel-1        & VV & 5 $\times$ 20 &           & DEM, InSAR coherence    & \\ \hline

\cite{aristizabal2020high}  &Sentinel-1 & VV, VH & 10  & Flood      &SAR intensity, HAND       & 5\\ \hline
 
\cite{chini2020role}  &Sentinel-1 & VV, VH & 20  & FB, FO       &SAR intensity, InSAR coherence     &5 \\ \hline
 
\multirow{2}{*}{\cite{ohki2020study}}  &\multirow{2}{*}{ALOS-2/PALSAR-2}  & \multirow{2}{*}{HH} & \multirow{2}{*}{3}  & \multirow{2}{*}{FB, FO}        &SAR amplitude, InSAR coherence, land cover/use map     & \multirow{2}{*}{5}  \\
                     &                         &          &  &            &  quasi-realtime flood fraction data  &  \\   \hline

\multirow{2}{*}{\cite{ohki2020automated}}   &\multirow{2}{*}{ALOS-2/PALSAR-2}  & \multirow{2}{*}{HH} & \multirow{2}{*}{3}  & \multirow{2}{*}{FB, FO}       &SAR amplitude, InSAR coherence, land cover/use map     &\multirow{2}{*}{5} \\
                  &          &  &   &         & quasi-realtime flood fraction data    & \\  \hline
 
\cite{ohki2020flood}  &ALOS-2/PALSAR-2  & HH & 10  & FB     &InSAR phase    &5\\ \hline

\cite{moya2020learning}  & Sentinel-1 &  VV & 14  & FB      &InSAR coherence    & 5 \\ \hline

\cite{liu2021inundation}  & Sentinel-1  & VV, VH & 10  & FB, FO      &SAR intensity, building footprint  &5 \\ \hline
 
\multirow{2}{*}{\cite{mason2021improving}}  & TerraSAR-X,  & \multirow{2}{*}{HH} & \multirow{2}{*}{3}  & \multirow{2}{*}{FB}   &SAR intensity, FRP maps,     &\multirow{2}{*}{5} \\
                   &  COSMO-SkyMed         &  &   &            & dynamic model flood inundation &   \\    \hline  
                    
\cite{mason2021floodwater}  & Sentinel-1& VV & 20  & FB     &SAR intensity, WorldDEM, WSF   &5\\ \hline
 
\cite{pulvirenti2021insar}  & Sentinel-1  & VV & 15  & FB     &InSAR coherence, InSAR phase  &5  \\ \hline
 
\cite{zhang2021urban}  & Sentinel-1  & VV, VH & 5 $\times$ 20  & FB      &SAR intensity, InSAR coherence  &5 \\ \hline
 
\cite{zhao2022urban}  &Sentinel-1& VV, VH & 20  & FB, FO      &SAR intensity, InSAR coherence   &7 \\ \hline
 
\cite{pelich2022mapping}  & Sentinel-1  & VV, VH & 10  & FB     &SAR intensity, InSAR coherence      &5 \\ \hline
 
\cite{bioresita2022integrating}  & Sentinel-1  & VH & -  & FB, FO      &SAR intensity, InSAR coherence, water mask     & 5 \\ \hline
 
\multirow{2}{*}{\cite{baghermanesh2022urban}}  & \multirow{2}{*}{TerraSAR-X} &  \multirow{2}{*}{HH} & \multirow{2}{*}{3}  & \multirow{2}{*}{FB}    &SAR intensity, InSAR coherence, InSAR phase,     &\multirow{2}{*}{5} \\
           &       &    &  &               &  PolSAR boxcar filter images, reflectivity maps  &    \\ \hline
 
\cite{gokon2023detecting}  & ALOS-2 & - & 3  & FB      &SAR intensity    &5 \\ \hline

\cite{yang2023promoting}  & Sentienel-1  & VV, VH & -  & Flood      &SAR intensity, DEM, incidence angle, land cover map   &5 \\ 

\hline
\bottomrule[2pt]

\end{longtable} 

\end{landscape}
\end{onecolumn}

\twocolumn[]


\section{Characteristics of urban floods}
\label{sec:Characteristics}

Urban flood scenarios vary significantly due to complex environmental factors. Building density, height, orientation, surrounding vegetation, and water depth all play a role. Additionally, SAR sensor characteristics, including polarization, line of sight (LoS), incidence angle, wavelength, and spatial resolution, can influence how floodwater appears in SAR images. The SAR backscatter can be significantly influenced by all these factors. To effectively map urban floods using SAR data, it is essential to identify the most appropriate sensor characteristics. This section provides a concise summary of SAR principles and relevant research findings on the relationship between sensor characteristics and environmental conditions for urban flood mapping. We will also complement this overview with current research findings on the relationship between sensor characteristics and environmental conditions that have a significant impact on the effectiveness of any urban flood mapping. Furthermore, the urban flood mapping studies included in Table~\ref{tab:list_paper} are categorised according to their input data, as shown in Table~\ref{tab:category_paper}.

\begin{table*}[]
\centering
\caption {\label{tab:category_paper} The categories of all urban flood mapping studies according to the input data.\\ (\textit{Proceedings are shown in italic})\centering} 
\begin{center}
\begin{NiceTabular}{l|l|l}
\toprule[2pt]
\hline
 Different Input Data & \# Studies & Reference\\
\hline

Intensity & 4 & \cite{pulvirenti2010analysis},\textit{\cite{kwak2017new}},\cite{lin2019urban},\cite{gokon2023detecting} \\ \hline
Coherence & 2& \textit{\cite{pulvirenti2019flood}},\cite{moya2020learning}\\
Phase & 2& \textit{\cite{ohki2019floodIGARSS}}, \cite{ohki2020flood}\\
Coherence + Phase & 1 & \cite{pulvirenti2021insar}\\
\hline

Intensity + LiDAR DEM/DSM & 5 & \cite{mason2010flood},\cite{mason2012near},\cite{mason2014detection},\cite{mason2018robust},\cite{olthof2020testing} \\
Intensity + WorldDEM + WSF	& 1 &	\cite{mason2021floodwater}\\
Intensity + HAND + Land Cover Maps &	1	&\cite{aristizabal2020high}\\
Intensity + DEM + Land Cover Maps	&1	&\cite{pierdicca2008integrating}\\
Intensity + DEM + Land Cover Maps + Incidence Angle	&1	& \textit{\cite{yang2023promoting}}\\
Intensity + Building Footprint&	2	& \textit{\cite{kwak2018effect}}, \cite{liu2021inundation}\\
Intensity + Land Cover Maps&	1&	\cite{liu2018detection},\cite{benoudjit2019novel}\\
Intensity + Hydraulic-derived Information	&4	& \textit{\cite{tanguy2014development}}, \cite{iervolino2015flooding},\cite{tanguy2017river}, \cite{mason2021improving}\\
Intensity + STRM Water Mask	&1	& \textit{
\cite{vanama2019change}}\\ \hline

\multirow{2}{*}{Intensity + Coherence}	& \multirow{2}{*}{15}	& \textit{\cite{chini2012analysis}}, \textit{\cite{chini2016sar}},\cite{pulvirenti2016use},\cite{chaabani2018flood},\textit{\cite{chini2018monitoring}},\textit{\cite{natsuaki2018band}},\cite{li2019urban}, \\
 & & \cite{chini2019probabilistic},\textit{\cite{chini2019sentinel}},\cite{li2019urbanRS},\cite{olthof2020testing},\textit{\cite{chini2020role}},\cite{zhang2021urban},\cite{zhao2022urban},\cite{pelich2022mapping}\\
Intensity + Coherence + Land Cover Maps	& 3&	\textit{\cite{pierdicca2014flood}},\cite{pulvirenti2015integration},\textit{\cite{pulvirenti2017detection}}\\
Intensity + Coherence + Water Mask	&1&	\textit{\cite{bioresita2022integrating}}\\
Intensity + Coherence + Building Polygon	&1&	\cite{ohki2019flood}\\
Intensity + Coherence + Land Cover Maps &	\multirow{2}{*}{2}&	\multirow{2}{*}{\textit{\cite{ohki2020study}}, \cite{ohki2020automated}}\\
+ Quasi-realtime Flood Fraction Data & & \\

\hline

Intensity + Coherence + Phase + PolaSAR	&	\multirow{2}{*}{1}&	\multirow{2}{*}{\cite{baghermanesh2022urban}}\\
Boxcar Filter Images + Reflectivity Maps & & \\
\hline
\bottomrule[2pt]

\end{NiceTabular}
\end{center}

\end{table*}

\subsection{SAR Intensity} \label{sec:intensity}

A two-dimensional complex SAR image represents the backscattering, which is composed of an intensity and a phase component \cite{moreira2013tutorial}. SAR intensity measures the strength of the returned signal, while SAR phase represents the distance between SAR sensors and ground objects. Generally, there are three common scattering mechanisms (Figure~\ref{fig:scatter}) in SAR imagery: surface scattering, volume scattering, and strong or hard target scattering. Figure~\ref{fig:scatter} illustrates that surface scattering happens when energy scatters or reflects from a surface boundary between two distinct yet uniform media. Volume scattering can happen in the layers of vegetation, the ground, dry sand or soil, and snow. In these places, reflection comes from many small scattering elements~\cite{richards2009remote}. Individual hard targets play a central role in scattering from urban features, and they can dominate a resolution cell with a high backscattering value. Hard targets are discrete, not distributed, and they scatter in a variety of ways. Two of the most common types of scattering mechanisms in urban areas are double-bounce scattering and scattering from a planar flat surface that is orientated perpendicularly or sub-perpendicularly to the sensor LoS, resulting in very high backscatter value, as typically seen in the case of a house's gable roof.

In urban flood areas, various scattering processes can occur, depending on the complexity of the buildings and surrounding environment. Typically, when open flood water reflects the signal in the specular direction, it produces low levels of backscatter and creates dark areas in SAR images. On the other hand, areas with buildings show high levels of backscatter and appear bright because of the double-bounce scattering that occurs between the building façades and the ground surfaces. Nevertheless, in urban areas experiencing flooding, there is a possibility of specular reflection taking place in completely submerged buildings, resulting in a decrease in SAR intensity. In partially flooded built-up regions, the double bounce effect is more pronounced due to the floodwater's smooth surface and higher dielectric constant compared to the non-flooded condition. The SAR intensity is increased by these two factors: the water level and the height of the building~\cite{li2019urban}. The visual representations of different urban flood scenarios can be referred to Figure~\ref{fig:scatter}.

A common approach for SAR-based urban flood mapping is to identify areas with a significant increase in backscatter intensity. This increase is often caused by the double-bounce effect between buildings and floodwater. To detect these changes, one typically needs a pre-flood image and a post-flood image \cite{pierdicca2008integrating,kwak2017new, pulvirenti2010analysis,riyanto2022three}. In some cases, it is beneficial to use multiple images covering the entire flood event \cite{lin2019urban}. Assuming that the SAR intensity is the key indicator for detecting urban floods and that its increase is expected when the flooding starts, some factors influencing the SAR intensity backscatter value, such as incidence angle, polarization, wavelength, urban structure, and surrounding vegetation, must be taken into account for a correct interpretation of backscattering behavior.

\begin{itemize}

\begin{figure}[]
    \centering
    \includegraphics[height=0.7\textheight]{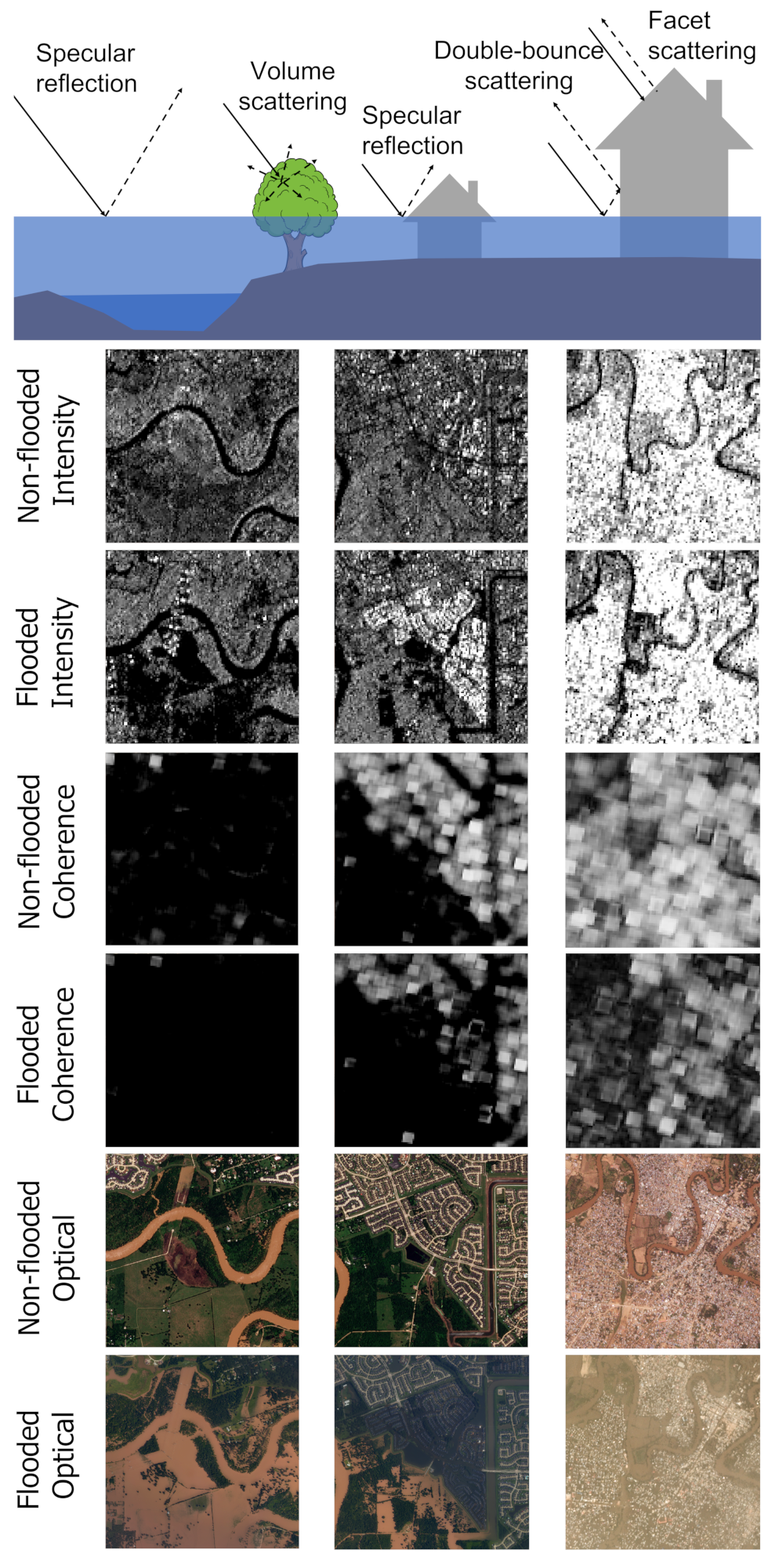}
    \caption{The scattering mechanisms in flooded urban areas \\ and examples of floods in different scenarios.}
\label{fig:scatter}
\end{figure}

\begin{figure*}[]
     \centering
     \includegraphics[width=0.7\textwidth]{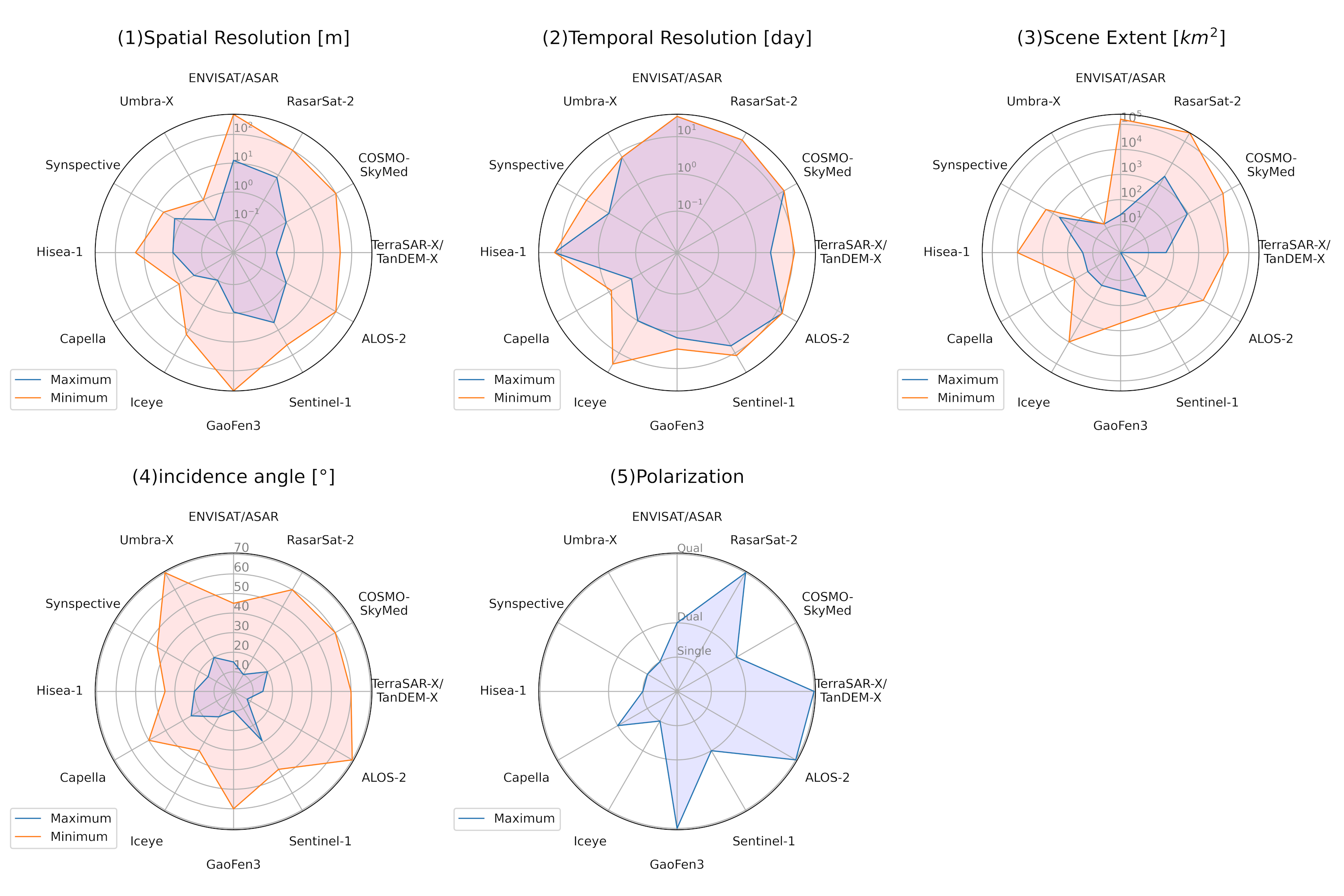}
     \hfill
\caption{Information about several SAR constellation which could be used in the flood mapping.}
\label{fig:common_SAR_info}
\end{figure*}

\item\textbf{The incidence angle} is defined as the angle between an imaginary line perpendicular to the Earth’s surface and radar signal. According to the specification of different SAR satellite sensors summarized in Figure~\ref{fig:common_SAR_info}, the incidence angle ranges between 8° and 70°. A large incidence angle is preferable for flood identification because it increases the contrast between land and water, especially in flooded open areas \cite{richards2009remote}, and it also improves ground range resolution. However, over-detection or/and under-detection may become severe since a large incidence angle contributes to a large shadowing, especially in urban areas with tall buildings. In general, a small incidence angle results in a smaller shadow area at the expense of a larger layover, which is an issue in the presence of tall structures, whereas a higher incidence angle results in a much smaller layover \cite{li2019urban, lin2019urban,mason2010flood}. Figure~\ref{fig:layover} illustrates these concepts using two buildings with height $h_1$ and $h_2$ separated by a flooded street $AD$. In this senario, the area $CD$ is in shadow due to the building's height while the area the $AB$ is affected by layover:

\begin{equation}
    CD=h_{2}tan\theta, AB=h_{1}cot\theta 
\end{equation}

However, shadow and layover regions may overlay narrow roads. To avoid this scenario, the minimum road width $w_{s}$ should satisfy the following condition as reported in \protect\cite{mason2010flood, soergel2003visibility}:

\begin{equation}
    w_{s} > AB + CD = h_{1}cot\theta  + h_{2}tan\theta
\end{equation}

Assuming that the buildings have the same height, i.e., $h_{1}=h_{2}=h$:
\begin{equation}
    w_{s} > h(cot\theta  + tan\theta)
\end{equation}

where the length of areas affected by a building’s layover is $h\times cot\theta$ and the length of areas affected by a building’s shadow is $h\times tan\theta$.

\begin{figure}[h!]
     \centering
     \begin{subfigure}[b]{0.95\linewidth} 
         \centering
         \includegraphics[width=\linewidth]{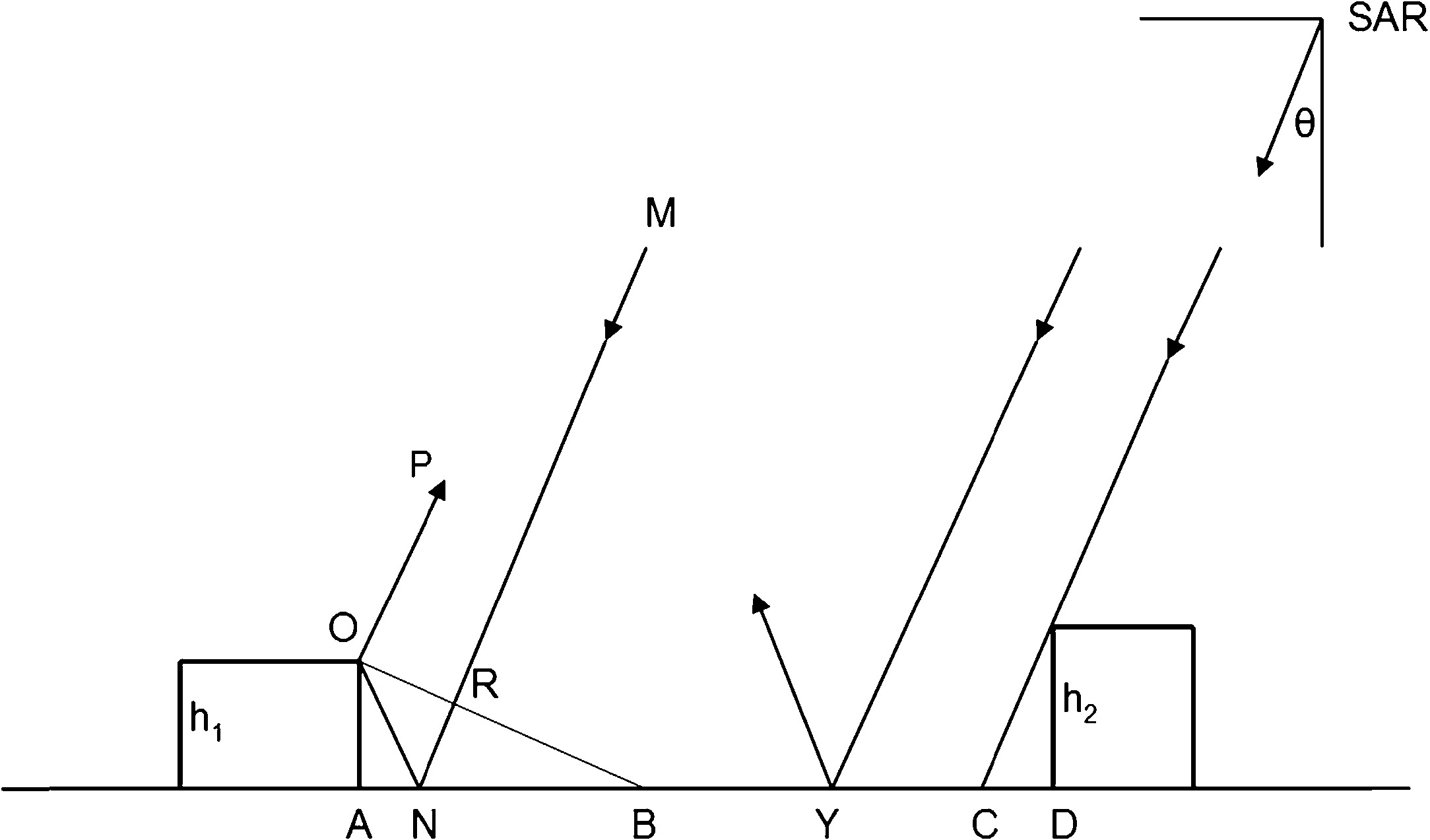}
         \caption{(AB) Layover and (CD) shadow regions in a (AD) flooded street between adjacent buildings of height $h_1$ and $h_2$ ($\theta$ = incidence angle) (adapted from \cite{mason2014detection})}
         \label{fig:layover1}
     \end{subfigure}
     \vfill
     \begin{subfigure}[b]{0.95\linewidth} 
         \centering
         \includegraphics[width=\linewidth]{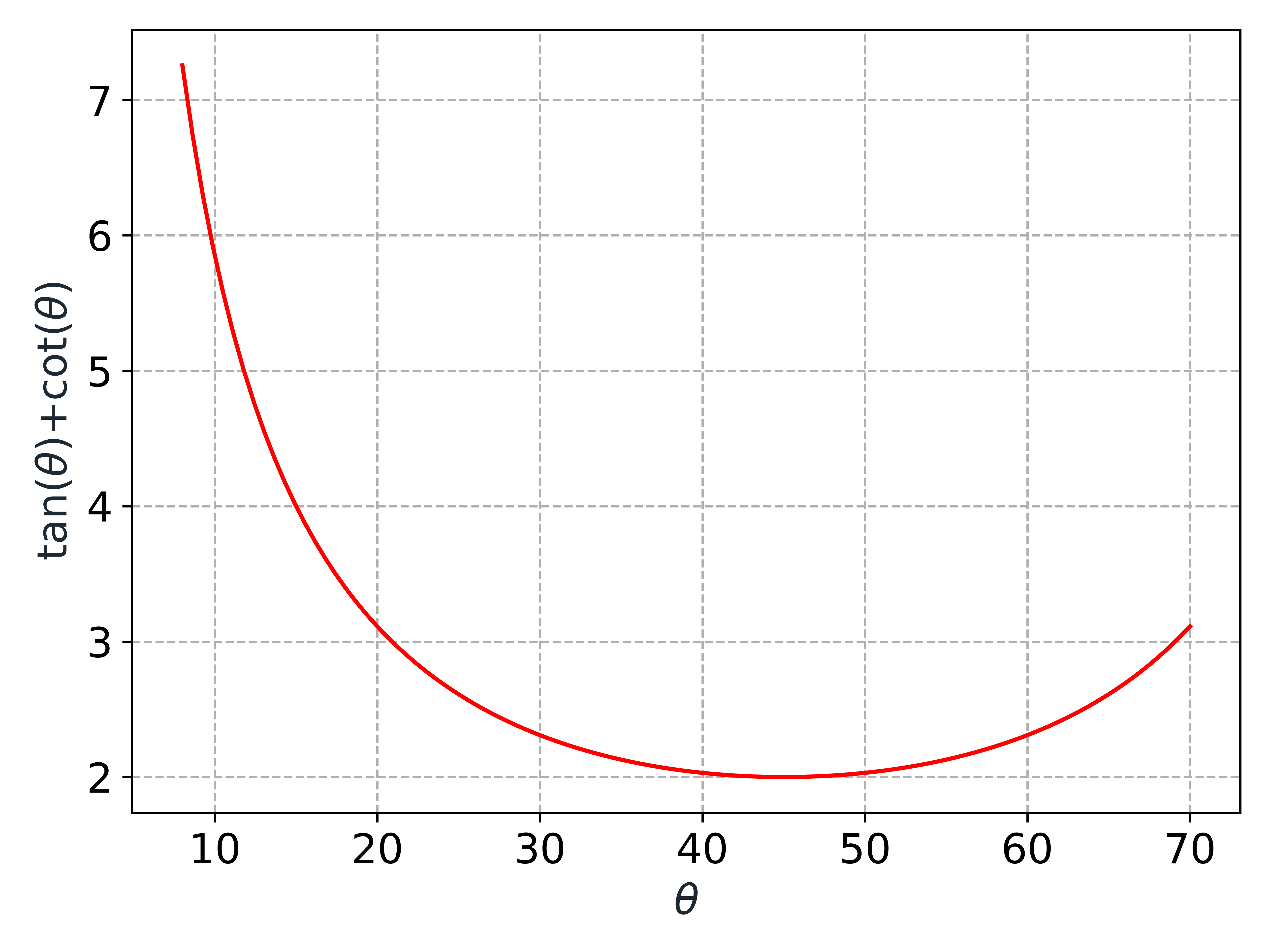}
         \caption{The relationship between the incidence angle and the minimal street width which can be sensed.}
         \label{fig:layover2}
     \end{subfigure}
\caption{Incidence angle effect in SAR intensity-based urban flood mapping.}
\label{fig:layover}
\end{figure}

Considering that the incidence angle of SAR sensors ranges between 8° and 70°, the minimum street width needed for flood detection can be calculated using the equation mentioned above. The optimal incidence angle for maximizing the number of detectable flooded pixels is around 45° (Figure~\ref{fig:layover2}). In this analysis, we only consider the simple scenario involving shadow and layover, while higher-order reflections from the ground and walls are neglected.

\item\textbf{Polarization} describes the orientation of the plane in which a propagating signal oscillates, encompassing linear, circle and elliptical polarization~\cite{long2015microwave}. Most spaceborne radar systems typically use linear polarization, with the most common basis polarizations being horizontal (H) and vertical (V) for a single channel. The signal of a SAR system can be transmitted and received using the same, i.e., co-polarized (VV or HH), and different polarizations, i.e., cross-polarized (VH or HV). Some advanced SAR systems are able to transmit and receive the signal in a dual-polarization mode (HH and VV, HH and HV, VV and VH) and in a quad-polarization mode (VV, HH, VH, HV). Dual and quad polarimetric sensors are able to identify different scattering mechanisms, which can be used to better characterise different land cover types \cite{tsyganskaya2018sar}. When mapping floods in open areas, horizontal-horizontal (HH) polarization is often preferred because it is less sensitive to wind conditions compared to vertical-vertical (VV) polarization, which is more sensitive to rough surfaces. Cross-polarization is primarily sensitive to volume scattering, making it less effective for urban applications where the double-bounce scattering process, which co-polarization measurements can clearly reveal, tends to be the most relevant scattering type \cite{meyer2019spaceborne}. Thus, co-polarized SAR data is always used in identifying flooded urban areas via identifying the backscatter increase from the strong double-bounce effect between floodwater and the building’s façades. The studies conducted by \cite{mason2023toward, zhang2021urban} demonstrated the superior capability of VV over VH in effectively discerning flooded from unflooded urban areas using intensity from dihedral scatterers. Nevertheless, according to previous studies \cite{chini2020role,li2019urban,zhao2022urban,pelich2022mapping},  co-polarization and cross-polarization are both needed to generate accurate urban flood maps since other scattering mechanisms such as multi-bounce scattering can occur and depend on building orientation with respect to the SAR sensor’s LoS.

\item\textbf{Wavelength} is another key factor to consider. During a flood event with heavy precipitation, atmospheric attenuation of the radar signal becomes noticeable. It has been reported that the amount of absorption and scatter of the radar signal due to water drops is higher at higher frequencies \cite{pierdicca2018flood}. Thus, the X-band SAR image can observe the high attenuation from heavy rain, showing a very low backscatter similar to floodwater \cite{pulvirenti2014discrimination}. Lower-frequency SAR signals, such as those in the C and L bands, are less affected by this issue. Additionally, wavelength plays a crucial role in determining the penetration ability of SAR signals into different mediums~\cite{meyer2019spaceborne}. 

For example, the X-band SAR signal has very limited capability to penetrate into broadleaf vegetation, and thus mostly interacts with leaves at the top of the tree canopy. With very high resolution X-band sensors, the vegetation can even give rise to shadow effects, a potential source of over detection of floodwater. On the other hand, the L-band signal achieves greater penetration into vegetation and allows for more interaction between the radar signal and large branches and tree trunks. Therefore, when it comes to flooded urban areas with vegetation, X-band SAR data struggles to identify floodwater due to vegetation blockage, unlike the L-band SAR signal. Furthermore, the spatial resolution of the SAR signal relates to the wavelength. This connection arises due to the inverse relationship between the slant range spatial resolution and the system bandwidth, which is inversely proportional to the wavelength while the azimuth resolution is directly influenced by the length of the sensor’s antenna \cite{moreira2013tutorial}. Therefore, longer wavelength corresponds to smaller bandwidth and subsequently lower slant range spatial resolution. Moreover, the backscatter is also dependent on the surface roughness with respect to the sensor wavelength; the higher the roughness, the higher is the backscattering value~\cite{long2015microwave}.

\item \textbf{Characteristics of urban structures} such as building density, building orientation, building's roof shape can also influence the floodwater detection in urban areas. This interference is caused by the density of nearby buildings above a certain height, which contributes to multiple radar reflections, especially when it comes to extracting urban water bodies from 10m spatial resolution Sentinel-1 intensity data. \cite{kwak2018effect} studied the effect of building orientation in urban flood mapping, emphasizing that the building's orientation more strongly influenced the double-bounce effect from a building façade than the building height, roof, and shape in the pilot areas using single-polarized (i.e., HH) ALOS-2 amplitudes images. One simulation experiment carried out by \cite{pulvirenti2016use} indicates that the SAR intensity drops from 11.5 dB to around ~3.5 dB once the building orientation angle (i.e., the angle between the orientation of the wall and the azimuth direction) increases from 0° to larger than 5-10°. Although the building’s orientation influences the SAR intensity, no impact has been found on InSAR coherence. Additionally, the building's roof plays a crucial role when considering InSAR coherence information in urban flood mapping. Coherence change detection methods primarily capture changes around the building façades caused by floodwaters, rather than changes on the roofs themselves. As a result, there is a risk of under-detection in flooded buildings with large, flat roofs that have not been completely submerged \cite{zhao2022urban}.  
 
\item\textbf{Urban vegetation} is also a potential source for under-detection. The presence of vegetation near buildings can cause SAR intensity to either increase or remain unchanged during flood events, depending on the structure and density of the vegetation in relation to the radar wavelength and polarization \cite{lin2019urban}. Furthermore, \cite{amitrano2016urban} concluded that the surrounding dense vegetation may hinder flooded urban mapping because it reduces the coherence of built-up areas even in the absence of floodwater, a phenomenon further confirmed in \cite{li2019urban,zhao2022urban}. 

\end{itemize}

\subsection{InSAR Coherence and phase} \label{sec:coherence_and_phase}

Another crucial element in SAR-based urban flood mapping is InSAR coherence, which can complement intensity-only data that may not always suffice to detect the presence of floodwater. Intensity is typically used under the assumption that floodwater in urban areas causes a decrease in backscattering in fully submerged areas and an increase in the double-bounce effect for partially submerged buildings. However, due to varying building orientations relative to the sensor's LoS, the significance of double-bounce scattering between floodwater and building façades may be minimal. To mitigate under-detection caused by the limitations of intensity data alone, InSAR coherence is incorporated. It measures the pixel-wise magnitude of the complex correlation between two SAR images, considering both amplitude and phase components. Built-up environments generally exhibit a steady and strong InSAR coherence, making them stable targets over time \cite{chini2018monitoring}. Therefore, observing a reduction in InSAR coherence around structures can facilitate the detection of urban floods \cite{chini2019sentinel}. InSAR coherence approaches utilise pre-event coherence maps, obtained from two images captured prior to the flood, and co-event coherence maps, generated from one image  acquired before the flood and one image during the flood \cite{moya2020learning}. Another approach involves analysing the coherence time series of stable scatterers, as described by \cite{pulvirenti2019flood}. Many studies that use thresholding classify one pixel as flooded when the drop-off of its InSAR coherence is greater than 0.3 \cite{chini2019sentinel,natsuaki2018band,ohki2022experiment}. However, other factors can also contribute to reduced coherence, even in urban areas not affected by flooding, particularly when heavy rainfall occurs over a wide region. \cite{pulvirenti2019flood} demonstrated that even in regions unaffected by floodwater, the co-event coherence may decrease. So, to correctly find pixels that are flooded in cities, you need to look at both the overall trend of coherence and the trend of coherence normalised by the average value of coherence over the stable scatterers spread across the entire scene. The relationship between InSAR coherence degradation and floodwater depth has also been investigated, although, according to \cite{ohki2022experiment}, no clear correlation has been found.

It is important to note that the InSAR coherence mentioned earlier is specifically related to the repeat-pass interferometric technique. This method relies on two SAR images captured from the same orbit at different times. The presence of floodwater leads to a reduction in coherence due to temporal decorrelation. However, behavior differs when using single-pass interferometric coherence. \cite{chaabani2018flood} presented an illustration of this, exploring bistatic InSAR coherence from the TanDEM-X mission for urban flood detection. Two satellites simultaneously obtain two SAR images to calculate the bistatic InSAR coherence, with negligible temporal decorrelation. Flooded buildings may exhibit high coherence values, sometimes even higher than non-flooded buildings, due to the stronger backscattering caused by the double-bounce effect, which enhances the signal-to-noise ratio~\cite{chaabani2018flood}.

\cite{ohki2019floodIGARSS} found that amplitude-based and open-water flood detection methods are more accurate than coherence-based urban flood mapping methods. For example, floods that occur in densely built areas may not be detected by InSAR coherence-based approaches because the high coherence remains even when the water level rises and the amplitude becomes stronger. Due to the abundance of point-like scatterers in urban areas,  \cite{ohki2020flood, ohki2019floodIGARSS} proposed utilising the phase component as an alternative to coherence. These studies utilised L-band SAR data acquired in HH polarisation from the ALOS mission. The strategy is based on the observation that high amplitude signals in a small target become dominant in the coherence calculation, while the phase of the pixels offers a small contribution, resulting in high coherence even when there are changes on the ground. Therefore, the focus should be on phase rather than amplitude for monitoring floods in urban areas. A recent study conducted an experiment that employed multi-temporal analysis of persistent scatterers (PS), using both the InSAR coherence of PS and the phase component of SAR data \cite{pulvirenti2021insar}. The study examined the consistency and alterations in the phase of stable scatterers in urban settings, emphasizing that the integration of these two characteristics enhanced the accuracy of the classification map.

Several parameters, including the spatial and temporal baseline of the involved SAR data and the configuration of the InSAR coherence calculation, i.e., computation window size, can influence InSAR coherence.

\begin{itemize}

\item \textbf{Spatial and Temporal Baseline:} In theory, longer spatial and temporal baselines result in poorer InSAR coherence \cite{moreira2013tutorial}. However, in SAR-based urban flood mapping, the temporal baseline has minimal impact on InSAR coherence because cities typically remain stable targets over time, unless there is significant vegetation, which causes temporal decorrelation. Instead, the spatial baseline plays a more dominant role in affecting coherence, especially with shorter wavelengths, as spatial baseline decorrelation is inversely proportional to wavelength \cite{zebker1992decorrelation}. To ensure consistent comparisons between pre- and co-event coherence maps, both maps must have the same temporal and spatial baselines to avoid bias from baseline difference.

\item \textbf{Window size:} When computing coherence, the size of the computation window is crucial. A larger window that includes elements like trees, shadows, or roads along with buildings may result in a coherence map with lower resolution and reduced coherence values~\cite{amitrano2016urban}; on the other hand, a smaller window could produce a higher spatial resolution coherence with higher noise~\cite{touzi1999coherence}. Currently, the most commonly used window size for urban flood mapping is 5×5 for very high-resolution SAR data \cite{li2019urban,schneider2006polarimetric} and 9×9 for high-resolution SAR data \cite{chini2019sentinel,pelich2022mapping,zhao2022urban}.

\end{itemize}

\subsection{SAR intensity and InSAR coherence} \label{sec:intensity_and_phase}

All previously described intensity-based approaches rely on the assumption that the presence of floodwater leads to an increase in backscatter due to the double bounce effects. Nevertheless, that might not hold true in an intricate urban setting. Specifically, it applies best in optimal conditions, such as when a building is in an isolated location with a uniform ground surface, and its façade is aligned parallel to the direction of the SAR flight~\cite{chini2016sar}. When buildings are surrounded by trees or other structures, and their façades are not parallel to the direction of the SAR flight, the backscatter patterns in both flooded and non-flooded situations can be similar, as described in Section~\ref{sec:intensity}. However, the coherence-only approaches are less precise since the coherence map has a relatively low spatial resolution, as stated in Section~\ref{sec:coherence_and_phase}. Consequently, InSAR coherence and intensity are often combined in order to address the limitations of both intensity-based and coherence-based approaches for urban flood mapping. Table~\ref{tab:category_paper} indicates that out of the 49 studies, 15 of them examined urban floods by utilising both SAR intensity and InSAR coherence simultaneously. These studies employed the same methods to identify areas affected by flooding: 1) There is a drop in coherence when comparing the co-event coherence to the pre-event coherence; 2) There is a change in backscatter, either an increase or decrease, when comparing the intensity after the event to the intensity before the event. Out of the 49 research studies that utilised intensity-based approaches, 17 of them incorporated other data such as land cover maps, water masks, building polygons, and hydraulic information. These studies combined such data with SAR intensity and InSAR coherence to generate accurate extent maps of urban floods. In order to enhance the accuracy of the extent maps, \cite{baghermanesh2022urban} used four SAR characteristics: SAR intensity, InSAR coherence, InSAR phase, and Polarimetric SAR features. In addition, they utilised supplementary data such as SAR simulated data, topographic data, and land cover maps.

\subsection{Other information sources} \label{sec:aux}

Although SAR data can be valuable for identifying water in urban areas, there are certain regions where it alone is insufficient for mapping the full extent of floowater. These regions include areas with shadows and layovers. In their studies, \cite{mason2018robust,mason2014detection,mason2012near,mason2010flood} utilised a high-resolution LiDAR DSM and SAR simulator to initially delineate the shadow and layover pixels. The authors then combined topographic data from the LiDAR DEM with insights from adjacent inundated bare soils to identify floodwater in areas where SAR data alone was insufficient~\cite{olthof2020testing}. \cite{aristizabal2020high} employed the Height Above Nearest Drainage (HAND) as a supplementary data source for the detection of inundated urban areas, akin to DEM/DSM. When comparing the detailed building information obtained from very high-resolution LiDAR DEMs/DSMs to that obtained from HAND and low-resolution DEMs (such as 12 m WorldDEM), it is evident that the latter two do not provide the same level of detail. Urban flood mapping combines building data from different sources, such as land cover maps and the World Settlement Footprint (WSF), with intensity and terrain information, as documented by various researchers \cite{aristizabal2020high,mason2021floodwater,pierdicca2008integrating,benoudjit2019novel,kwak2018effect,liu2021inundation,liu2018detection}. To be more precise, the distinction between urban and non-urban areas is made using information obtained from building footprints. Subsequently, SAR intensity is utilised to differentiate between urban areas that are flooded and those that are not inundated, as explained in Section~\ref{sec:intensity}.\cite{vanama2019change} illustrate the application of SRTM water masks for distinguishing between persistent water bodies and floodwater. Enhancing urban flood mapping \cite{iervolino2015flooding,mason2021improving,tanguy2017river,tanguy2014development} can be achieved by including SAR intensity data in conjunction with other hydrological and hydraulic model-derived data, including gauge information, flood return period maps, and probabilistic flood maps.

\section{SAR-based urban flood mapping approaches}
\label{sec:method_section}

We provide a thorough analysis of existing techniques for mapping urban floods, taking into account their strengths and weaknesses. This encompasses the methods of visual inspection (Section~\ref{sec:Visual}), rule-based analysis (Section~\ref{sec:Rulebased}), electromagnetic model-based techniques (Section~\ref{sec:model}), and machine learning algorithms (Section~\ref{sec:ML}). Table~\ref{tab:method_catogery}  presents the categorization of all works according to the methodology employed. It is important to understand that the number of approaches used in a study may be larger than the number of studies selected, as a single study can employ multiple strategies. Moreover, included below is an elucidation of the four categories to assist readers in selecting the most appropriate approach for practical implementations.

\begin{table*}[ht!]
\centering

\caption {\label{tab:method_catogery} Information of four categories methods among the selected 49 studies. \\ (\textit{Proceedings are shown in italic}) \centering} 
\begin{center}
\begin{NiceTabular}{ccc|c|c}
\toprule[2pt]
\hline
\multicolumn{3}{c}{\multirow{2}{*}{Categories}}                                        & \# Use & \multirow{2}{*}{Reference}         \\
\multicolumn{3}{c}{}                                                                   & cases &                             \\ 
\hline
\multicolumn{3}{c}{Visual inspection}                                                  & 3         & \textit{\cite{chini2012analysis}},\textit{\cite{chini2016sar}}                     \\ \hline
\multicolumn{1}{c|}{\multirow{7}{*}{Rule-based approach}} &
  \multicolumn{2}{c|}{Fuzzy-logic based approaches} &
  2 &
  \cite{pierdicca2008integrating},\cite{tanguy2017river} \\ \cline{2-5} 
\multicolumn{1}{c|}{} & \multicolumn{1}{c|}{Region growing} & Manually selected seeds   & 5         & \cite{mason2010flood},\cite{mason2012near},\cite{mason2014detection},\cite{mason2018robust},\cite{olthof2020testing}                 \\ \cline{3-5} 
\multicolumn{1}{c|}{} & \multicolumn{1}{c|}{based approach} & Rule-based selected seeds & 4         & \cite{pierdicca2014flood},\cite{pulvirenti2015integration},\cite{pulvirenti2016use},\textit{\cite{pulvirenti2017detection}}                    \\ \cline{2-5} 
\multicolumn{1}{c|}{} & \multicolumn{2}{c|}{Decision tree-based approaches}             & 2         & \cite{pulvirenti2010analysis},\textit{\cite{natsuaki2018band}},\cite{pulvirenti2021insar}                        \\ \cline{2-5} 
\multicolumn{1}{c|}{} &
  \multicolumn{1}{c|}{\multirow{3}{*}{Thresholding}} &
  \multirow{2}{*}{Manual threshold} &
  \multirow{2}{*}{17} &
  \textit{\cite{tanguy2014development,kwak2017new,kwak2018effect,chini2018monitoring,pulvirenti2019flood,chini2019sentinel}}, \cite{liu2018detection,ohki2019flood},\\
\multicolumn{1}{c|}{} & \multicolumn{1}{c|}{}               &                           &           & \textit{\cite{ohki2019floodIGARSS,chini2020role,bioresita2022integrating}},\cite{olthof2020testing,ohki2020flood,liu2021inundation,mason2021improving,mason2021floodwater,zhang2021urban} \\ \cline{3-5} 
\multicolumn{1}{c|}{} & \multicolumn{1}{c|}{}               & Automatic threshold        & 2         & \textit{\cite{vanama2019change}},\cite{pelich2022mapping}                       \\ \hline
\multicolumn{3}{c}{Electromagnetic model-based approaches}                             & 1         & \cite{iervolino2015flooding}                         \\ \hline
\multicolumn{1}{c|}{\multirow{3}{*}{Machine Learning}} &
  \multicolumn{2}{c|}{Bayesian inference-based approaches} &
  5 &
  \cite{lin2019urban},\cite{chini2019probabilistic},\cite{li2019urbanRS},\textit{\cite{ohki2020study}},\cite{ohki2020automated} \\ \cline{2-5} 
\multicolumn{1}{c|}{} & \multicolumn{2}{c|}{Machine learning classifier}                & 6         & \cite{chaabani2018flood}, \cite{benoudjit2019novel}, \cite{aristizabal2020high},\cite{moya2020learning}, \cite{baghermanesh2022urban}, \cite{gokon2023detecting}       \\ \cline{2-5} 
\multicolumn{1}{c|}{} & \multicolumn{2}{c|}{Deep Learning}                              & 3         & \cite{li2019urban},\cite{zhao2022urban},\textit{\cite{yang2023promoting}}                    \\ \hline
\bottomrule[2pt]
\end{NiceTabular}
\end{center}
\end{table*}

\subsection{Visual inspection}\label{sec:Visual}

\begin{figure}[htb!]
     \centering
     \includegraphics[width=\linewidth]{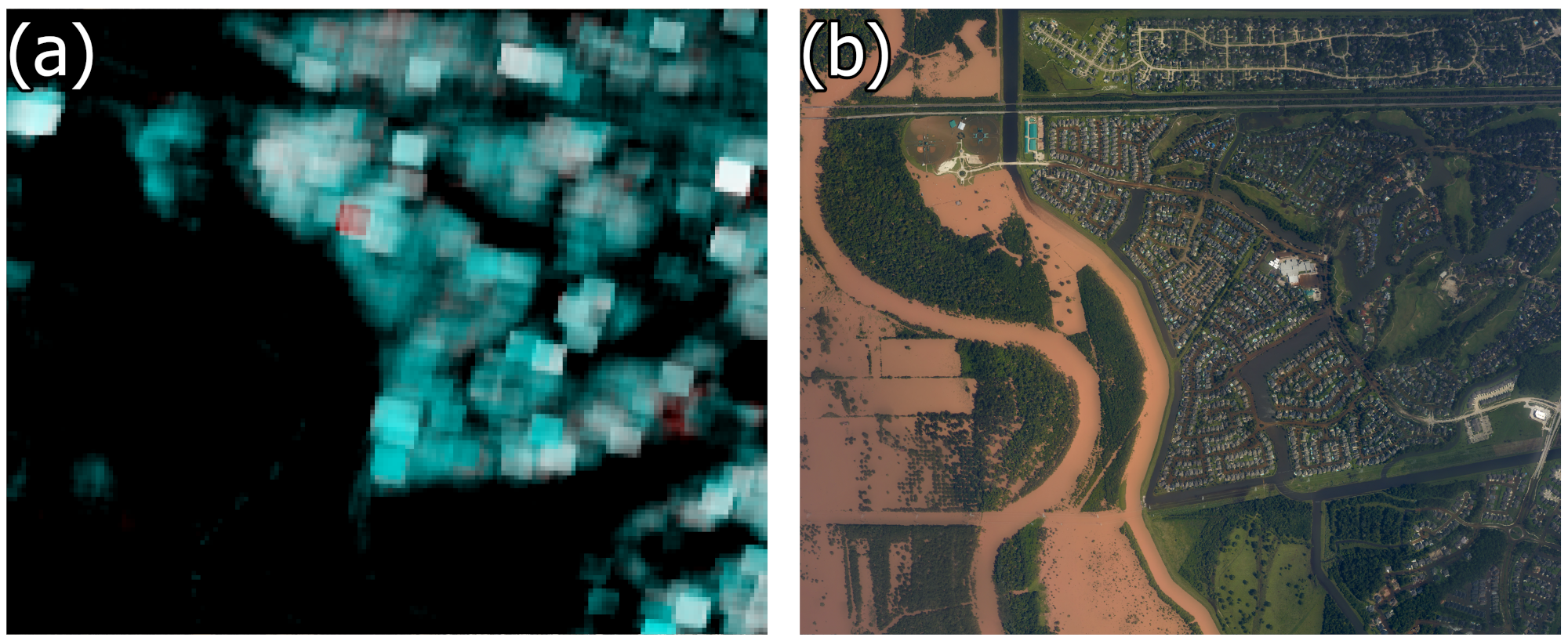}
\caption{Example of visual inspection-based urban flood mapping: (a) RGB composition of pre- and co-event InSAR coherence image in the Houston areas (R=co-event image, G=B=pre-event image); (b) optical image acquired after the flood event provided by NOAA.}
\label{fig:visual_exam}
\end{figure}

The conventional and direct approaches in SAR-based urban flood mapping entail the utilisation of visual inspection techniques \cite{chini2016sar,chini2012analysis}. These methods depend on the $RGB$ visualisation of coherence prior to and following an event. It is assumed that the loss of coherence is common during urban flood occurrences, as stated in Section~\ref{sec:coherence_and_phase}. Figures~\ref{fig:visual_exam} (a) and (b) depict two Sentinel-1 images taken over a flooded urban area in Harris County, US. Figure~\ref{fig:visual_exam} (a) shows the co-event coherence by using Sentinel-1 data obtained on August 24 and 30, 2017, and the pre-event coherence by employing images obtained on August 18 and 24, 2017. Figure~\ref{fig:visual_exam} (b) represents the optical UAV data collected on August 31, 2017, with a spatial resolution of 35 cm. By visually examining Figure~\ref{fig:visual_exam} (a) and detecting the cyan-colored areas indicating reduced coherence, one can promptly identify the flooding in built-up regions in Figure~\ref{fig:visual_exam} (b) through visual inspection. Visual inspection approaches have limitations as they can only offer qualitative information and lack the capability to draw flood extent maps. Consequently, more advanced methods have been developed to generate accurate flood extent maps.

\subsection{Rule-based approaches}\label{sec:Rulebased}

\subsubsection{Fuzzy logic based methods}

Two studies \cite{pierdicca2008integrating,tanguy2017river} used SAR intensity and fuzzy logic to delineate flooding in urban and non-urban areas, in combination with different types of data, like land cover maps, topography maps, and flood hazard maps. \cite{pierdicca2008integrating} focused on flooded, built-up areas with strong double-bounce effects, integrating topographic information and land cover maps with SAR intensity. Urban floodwater effectively increased backscatter, while open floods caused a decrease. Thus, the situation where high floodwater depth in urban areas leads to decreased SAR backscatter has yet to be considered. Moreover, they utilised topography data to reclassify sparsely distributed pixels, initially identified as non-flooded by SAR data, as flooded ones when their surroundings show higher elevation values. In contrast, \cite{tanguy2017river} first detected 'open water' in urban areas using SAR intensity data and refined their results by including a flood hazard scenario. This method did not consider detecting flooded urban areas with increased backscatter due to double-bounce effects. While fuzzy logic-based methods are capable of considering multiple variables for classification and making appropriate decisions, they are not easily scalable due to the involvement of fuzzy sets and fuzzy thresholds, which are primarily fixed based on experience.

\subsubsection{Region growing based methods}

Region growing is a subcategory of methods that involves examining the neighbouring pixels of initial seeds to determine their suitability for inclusion in the classified region. Table~\ref{tab:method_catogery} indicates that among the 49 studies surveyed, nine employed region-growing techniques. We can further categorise these methods into two groups: manual seed selection \cite{mason2018robust,mason2012near,mason2010flood,olthof2020testing} and seed selection based on set rules \cite{giustarini2013change,pierdicca2014flood,pulvirenti2017detection,pulvirenti2016use,pulvirenti2015integration}. \cite{mason2018robust,mason2014detection,mason2012near,mason2010flood} explored the use of region-growing to delineate urban flood boundaries by incorporating backscatter, topography, and flood edge data from adjacent rural areas. This approach's fundamental principle is to capitalise on backscatter and water height thresholds while having prior knowledge of flood extent in nearby rural regions. The region-growing process involves the selection of manual seed points with low backscatter values in urban regions. These seed points act as the initial starting locations for the region-growing algorithm. Subsequently, the algorithm progressively extends into unexplored areas by employing a water height threshold derived from a LiDAR DEM. The rationale behind this approach is that potentially flood-prone urban areas should not have substantially higher elevations than their neighbouring rural counterparts. Furthermore, they used a SAR simulator to mask out shadows and layovers in the SAR data. \cite{olthof2020testing} introduced a similar method, but it eliminates the need for a SAR simulator by disregarding shadows and layovers. The decision to omit these steps aims to expedite the flood extent mapping process, thereby delivering faster results. However, the manual selection of seeds in the region's growing process demands considerable expertise and is time-consuming. As a result, subsequent studies introduced predefined rules to determine seed points \cite{pierdicca2014flood,pulvirenti2016use,pulvirenti2015integration}. In particular, they picked seeds and set the boundary to stop the region's growth using a supervised analysis and the rules from Section~\ref{sec:Characteristics}, which include a drop in coherence and a change in intensity. The distinction between these studies lies in the variations in the rules used for the region-growing algorithm. \cite{pierdicca2014flood} considered pixels with a decrease in coherence and an increase or decrease in backscatter as flooded urban areas; \cite{pulvirenti2016use,pulvirenti2015integration} considered pixels with a decrease in coherence and an increase in backscatter as flooded urban areas. It is worth noting that the aforementioned studies used the CORINE land cover map to select urban flood seeds and identify urban areas. Instead of using additional land cover maps, \cite{pulvirenti2017detection} proposed a similar urban flood mapping method that derived the urban map by identifying stable scatterers through time series of coherence data.

\subsubsection{Decision tree-based approaches}

\begin{figure}[h]
     \centering
     \includegraphics[width=\linewidth]{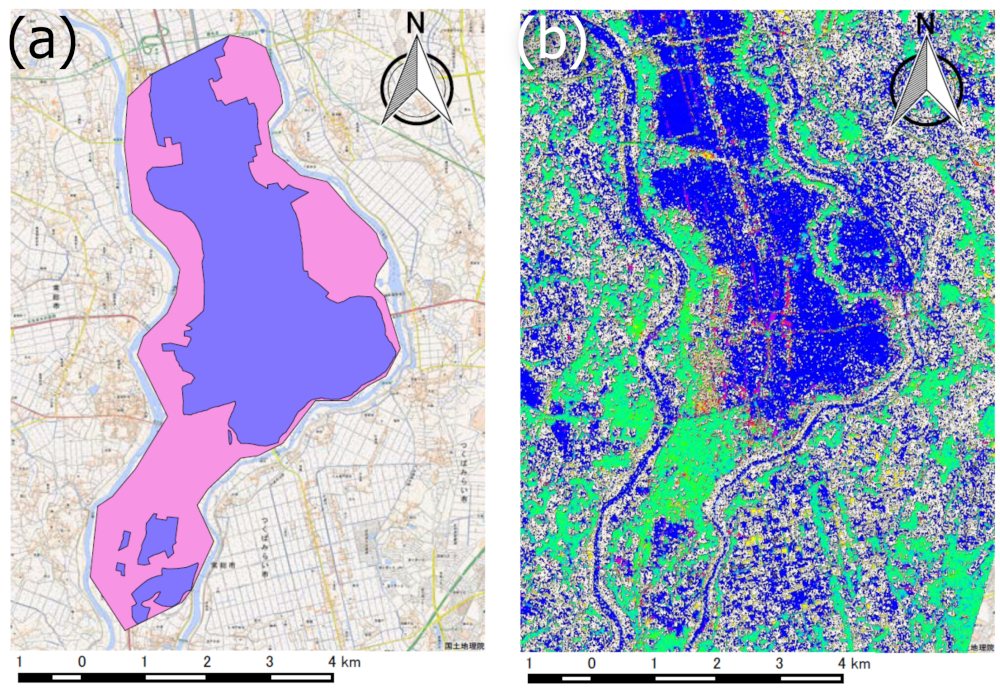}
\caption{Inundation maps generated using the decision tree-based flood detection scheme proposed in \cite{natsuaki2018band}: (a) The pink polygon highlights the evaluated region, while the blue polygon denotes inundated areas reported by the Ministry of Land, Infrastructure, Transport, and Tourism (MLIT), Japan. (b) Yellow represents moderately affected flood pixels in urban areas, and purple denotes severely flooded urban pixels.}
\label{fig:natsuaki2018paper}
\end{figure}

The inclusion of decision tree-based and rule-based techniques in urban flood mapping was motivated by their flexibility and robustness \cite{pulvirenti2021insar,pulvirenti2010analysis}. Once rules are established based on expert knowledge, these methods can deliver classification results more quickly than machine learning approaches. For example, \cite{pulvirenti2010analysis} identified flooded urban areas by setting a predefined threshold for decreases in SAR backscatter. Similarly, \cite{natsuaki2018band} employed a flood detection scheme comprising 5 rules: 1) Pixels exhibiting an intensity decrease greater than 6 $dB$ were classified as blue; 2) Pixels showing an intensity increase greater than 6 $dB$ were classified as yellow; 3) Pixels with a coherence drop-off larger than 0.3 were labeled red; 4) Pixels with a coherence increase higher than 0.6 were designated as green; 5) Pixels fulfilling both Rule 3) and Rule 4) were left unassigned, without any color designation. Then, the urban pixels moderately affected by the flood event were represented as yellow, while severely flooded urban pixels were depicted in purple based on visual inspection (Figure~\ref{fig:natsuaki2018paper}). More recently, \cite{pulvirenti2021insar} employed a logical $AND$ rule to categorise urban flood pixels and identified areas with flooded permanent scatters (PS). This rule considered modifications in coherence and the interferometric phase. Typically, region-growing, decision tree-based, and rule-based techniques rely on pre-existing information. Nevertheless, the documented precise rules and observed measurable limits offer essential knowledge and understanding from several study locations and data origins. This information can serve as a basis for our future research and decision-making processes.

\subsubsection{Thresholding}
\label{sec:Thresholding}
Table~\ref{tab:method_catogery} clearly shows that thresholding is the most commonly used method in urban flood mapping, as it allows for the direct identification of flooded areas by applying a specific threshold value. Out of the 49 research, 19 have utilised this method due to its simplicity and efficacy. It is remarkable that manual thresholding approaches continue to be the prevailing approach, given that 17 out of 19 thresholding studies manually chose their thresholds. \cite{liu2021inundation,liu2018detection} employed a threshold to identify buildings that were partially inundated. The SAR intensity threshold was determined by calculating the average intensity value over selected flooded sites and adding three standard deviations. \cite{kwak2018effect,kwak2017new} detected urban floodwater by analysing the correlation coefficient of pre- and post-event intensity images, in addition to employing SAR intensity alone. As elucidated in Section~\ref{sec:coherence_and_phase}, coherence plays a crucial role in flood detection. \cite{bioresita2022integrating} applied a coherence reduction threshold of 0.3 directly to the coherence data, as the presence of floodwater in urban areas typically reduces coherence. Similarly, \cite{chini2020role,chini2019sentinel,chini2018monitoring} identified urban areas susceptible to flooding by applying a thresholding technique to the pre- and co-event coherence data, using a building map as a reference mask to define the specific regions for coherence analysis. The SAR time series was utilised to automatically derive this map by analysing its multitemporal intensity and coherence. In contrast,~\cite{pulvirenti2019flood} used stable scatterer (StSc) urban pixels as the primary reference for urban flood mapping, rather than extracting the building map from SAR data. More precisely, they determined the inverse coefficient of variation (ICV) for every pixel within an urban area. The ICV is a measure of the relationship between the average amplitude of the radar backscatter and its variability over a period of time. Ultimately, pixels were categorised by their coherence value, which was then compared to the minimal coherence value of stable scatterers. However, InSAR coherence analysis alone was insufficient to detect some flood-affected urban neighborhoods, leading to the inclusion of phase information.~\cite{ohki2020flood,ohki2019flood} employed trial-and-error criteria to examine the intensity, coherence, and phase standard deviation using L-BAND ALOS PALSAR-2 data. They discovered that the phase standard deviation is a more precise measure than coherence. This is because the phase primarily concentrates on the geometric alterations in the scattering process, which are not influenced by the magnitude of backscattering from structures.

\cite{zhang2021urban} applied an empirical threshold of 2.40 to an urban flood index, defined as the ratio of backscatter coefficient increasing (BCI) to interferometric coherence decrease (ICD) in VV polarisation, rather than directly thresholding SAR data. This approach avoided the need for an additional land cover map by applying a random forest to the pre-flooding PolSAR images, which resulted in a land cover map that identified urban areas.~\cite{mason2021floodwater,mason2021improving} mapped urban floods using a combined water level threshold map, which integrated both model-based and SAR-derived waterline mapping. Following this combination, they applied thresholding to the combined maps, leveraging accurate elevation data to accurately detect inundated urban regions. The SAR-derived waterline map was made by combining nearby SAR-derived rural flood extents with a high-resolution LiDAR DSM, based on an assumption that water levels in cities shouldn't be much higher than those seen in the rural areas nearby.

Automatic thresholding is more effective, particularly for large-scale flood mapping, due to the strong correlation between the ground's dielectric constant and the incidence angle of the SAR sensors with the SAR backscatter. This indicates that it is required to make specific adjustments to the thresholds for each case \cite{chini2017hierarchical,zhao2021large}. \cite{vanama2019change} employed the Kapur algorithm for automatic thresholding in their 2019 study on the normalised change index (NCI) obtained from both the flood and pre-flood data in VV polarisation. The Kapur algorithm, proposed by \cite{kapur1985new}, determines the threshold that maximises the difference in entropy between the two groups. Classifying changes in backscatter due to urban floodwater without considering coherence often leads to an underestimation of the flood extent when relying solely on intensity. This approach fails to detect certain flooded pixels in densely populated urban areas where backscatter remains constant during the flooding event. Furthermore, the ability to distinguish between the two classes is determined by the distribution of classes within the image and the extent to which the two classes overlap. In order to address this constraint, \cite{chini2017hierarchical} proposed a hierarchical split-based approach (HSBA) that partitions the classes of a bimodal distribution by detecting image tiles where the two distributions may be fitted with higher accuracy and precision. \cite{pelich2022mapping} employed HSBA to detect the decrease in coherence by integrating this automatic adaptive thresholding method with a region-growing technique.

\subsection{Electromagnetic Model-based approaches}\label{sec:model}

In addition to the statistics-based classification methods discussed earlier, deterministic analysis provides another solution for urban flood mapping. The Kirchhoff scattering model is used to estimate the double-bounce effect arising from floodwater. This model characterizes electromagnetic scattering from surfaces featuring gentle undulations, where the average dimensions are significantly larger than the signal wavelength \cite{long2015microwave}. In the context of urban flood mapping, two distinct approximations of the Kirchhoff scattering model are utilized. The first is the Geometrical Optics (GO) model, while the second is the Physical Optics (PO) model. At each bounce/single scattering, GO or PO can be employed, depending on the wall or terrain roughness. More specifically, the GO is suitable when $ks=(2\pi s/\lambda) \geq 3$ while PO is suitable when $ks <3$. In this context, $s$ is rms height and $\lambda$ is the wavelength \cite{long2015microwave}. When it comes to the double bounce scattering in the urban environment, GO is used to evaluate the field reflected by the smooth wall towards the ground (first bounce) or the sensor (second or third bounce) while GO or PO (according to ground surface roughness) is used to evaluate the field scattered by the ground toward the wall (first or second bounce) or sensor (second bounce) \cite{franceschetti2002canonical}. Consequently, the double-bounce scattering from wall-to-ground reflection can be modelled using either the GO-GO model or the GO-PO model. Similarly, double bounce scattering from ground-to-wall adopts the same model, due to the equal path length taken by waves that initially hit the ground and then the wall, or vice versa \cite{franceschetti2002canonical}. While the GO-PO model proves more appropriate for lower ground roughness compared to the GO-GO model, it is noteworthy that the GO-GO model emerged as the more suitable approximation for scenarios ranging from normal to flooding conditions \cite{iervolino2015flooding,mason2014detection}. The rationale lies in the absence of significant alterations in ground roughness at the scale of normal and flooding conditions. This lack of change can be attributed to tall trees and dense vegetation near the building, effectively concealing any changes at the ground roughness level.

The advantage of the method proposed in \cite{iervolino2015flooding} is that the GO-GO model can be expressed by closed-form equations, enabling the straightforward inversion of different variables, such as the water level. However, due to the complexity of the model, numerous parameters (e.g., aspect angle, building height, complex dielectric constants of ground and wall, roughness of ground and water) must be assumed in advance. This introduces uncertainties related to prior knowledge and the model itself, affecting the accuracy of the final identified urban flood extent.

\subsection{Machine learning methods} \label{sec:ML}
The final category encompasses machine learning methods, comprising 14 studies out of 49 conducted investigations, which can be further divided into three subsets.

\subsubsection{Bayesian inference-based approaches}

Most machine learning methods in this category rely on Bayesian inference, where flooded pixels are classified based on the posterior probability of flood conditions using the Bayesian probabilistic model and different input data. Therefore, a simple Bayesian probabilistic model was adopted to obtain the posterior probability of flood states $F$:
\begin{equation} \label{eq:bayes_orig}
    p(F_i|x)=\frac{p(x|F_i )p(F_i )}{\sum_{j=1}^{k}(x|F_i)p(F_i)}
\end{equation}
	
where $x$ is the observed data while $F_i$ are different flood states.

\cite{chini2019probabilistic} introduced the post-event intensity, pre/post-event intensity changes and pre/co-event coherence changes to the Bayesian probabilistic model where the conditional probability calculated by the HSBA (HSBA was introduced in section~\ref{sec:Thresholding}). In this study, only two flood states were defined: $F_0$ was the non-flood state and $F_1$ was flood state. For each individual pixel, the estimation of the conditional probability of its inundation status, given its corresponding backscatter value, can be achieved as follows:

\begin{equation} \label{eq:bayes_chini_2019}
    \begin{gathered}
        p(F_1|x)=\frac{p(x|F_1)p(F_1)}{p(x|F_1)p(F_1)+ p(x|F_0)p(F_0)} \\
        with\ x\in \left \{  \sigma^0,\Delta\sigma^0, \Delta\gamma \right \}
    \end{gathered}
\end{equation}

where $p(F_1 )$ is the probability distribution of backscatter values of a flooded urban pixel, $p(F_0 )$ is the probability distribution of backscatter values of an unflooded urban pixel, $p(F_1 )$  and $p(F_0 )$ are fixed as 0.5. It should be noted that different SAR features were used for floodwater mapping on bare soils and urban areas, respectively. For bare soils, $x$ is SAR intensity $\sigma^0$ and changes of SAR intensity $\Delta\sigma^0$. For urban areas,  $x$ is the value from SAR intensity  $\sigma^0$, changes of SAR intensity  $\Delta\sigma^0$ and changes of InSAR coherence $\Delta\gamma$. The $x$ probability distributions of flooded pixels and unflooded pixels are assumed to follow Gaussian distributions, which can be parameterized using the identified pixels through HSBA. Thus, with all the known parameters in Equation~\ref{eq:bayes_chini_2019}, the probability of flooded pixels from different input features $\sigma^0$,  $\Delta\sigma^0$ and $\Delta\gamma$ can be calculated and combined to identify flooded urban and bare soil areas. 

Similarly, \cite{lin2019urban} estimated the conditional probability of normalized multi-temporal intensity data using the Levenberg-Marquardt algorithm for Bayesian inference, assuming that the peak value of the time-series of SAR intensity belongs to the inundated urban areas.  Specifically, this study utilized a multi-temporal SAR intensity approach, which was subjected to normalization utilizing the mean and standard deviation of non-flood epochs within the temporal series. This normalization was conducted through the exclusion of the flooding epoch for each pixel, achieved by applying z-score transformation in the temporal domain. As a result, the mean and standard deviation of the normalized intensity time-series for each non-flooded pixel were adjusted to 0 and 1, respectively. For flooded pixels, the mean and standard deviation of the normalized flooded pixels were calculated in the spatial domain as only one flooded epoch existed in this study. Assuming the distribution of flooded and unflooded areas follow Gaussian distributions and $p(F_0)=p(F_1 )=0.5$, each term in Equation~\ref{eq:bayes_chini_2019} can be calculated analytically and the conditional probability of flooded areas and non-flooded areas were finally estimated according to Equation~\ref{eq:bayes_chini_2019}. In this study, a probability cutoff value was defined as 0.5 in order to generate a binary flood map from the flood probability map.

However, establishing a direct causal relationship between flood state and observed SAR features (such as intensity and coherence) in real-world scenarios is challenging, particularly in urban areas where complex backscattering mechanisms arise due to diverse land covers. \cite{li2019urbanRS} proposed to decouple the hidden variable C between flood state and SAR features via the probabilistic graphic model. Therefore, Equation~\ref{eq:bayes_orig} can be converted into:

\begin{multline}
        p(F_1|\sigma^0,\rho) = \\
        \frac{\sum_{C_\rho}^{}p(\sigma^0|C_\rho)p(C_\rho|F_1) \sum_{C_\gamma}^{}p(\sigma^0|C_\gamma)p(C_\gamma|F_1)}{\sum_{F}^{}\sum_{C_\rho}^{}p(\sigma^0|C_\rho)p(C_\rho|F)\sum_{C_\gamma}^{}p(\sigma^0|C_\gamma)p(C_\gamma|F)p(F)} \\
        p(C_j|F)=\frac{p(F|C_j)p(C_j)}{\sum_{C_j}^{} p(F|C_j)p(C_j)}, with\ j \in \left \{ \sigma, \gamma \right \}
        \label{eq:bayes_Li_2019}
\end{multline}

where $p(F)$ and $p(F)$ can be calculated via the Bayes rule.

In Equation~\ref{eq:bayes_Li_2019}, $p(C_\sigma)$ and $p(C_\gamma)$ follow Gaussian distributions, whose mean and standard deviation can be estimated by a finite Gaussian Mixture Model (GMM). The non-informative prior probability was used that $p(F_0)=p(F_1)=0.5$. The $p(C_j)$ was the vector of weights of $K$ Gaussian component, while $K$ was selected via the Bayesian information criterion (BIC). The $p(F|C_j)$ was estimated based on the variation between the average of pre-event series and co-event acquisition for each component centroid via the sigmoid function. Therefore, the joint probability distribution integrating time-series intensity and coherence can be parameterized over a fixed set of random variables in order to detect flooded pixels.

\cite{ohki2020study} also adopted the Bayesian probabilistic model in their study where four states instead of 2 or 3 were assumed. Thus, in Equation~\ref{eq:bayes_orig}, $k=0,1,2,3$ and three SAR features was considered: $F_0$ is the non-flood state, $F_1$ is the unclassified areas, $F_2$ is the open-water flood, $F_3$ is the flooded building; $x \in \{\sigma_{co}^0,(\sigma_1^{pre},…,\sigma_4^{pre})  ,\gamma_{co} - \gamma_{co} \}$. Similarly, $p(F_i)$ was assumed as Gaussian distribution and determined using the empirical parameters. The prior probability $p(F_i)$ was defined different from above mentioned studies:

\begin{equation} \label{eq:Ohki_2020}
    \begin{gathered}
        p(F_0)=p(F_1)=0.5(1-p_f) \\
        p(F_2)=p(F_3)=0.5p_f \\
        p_f(y)=\frac{1}{1+e^{-\frac{1}{\sigma}(y-\mu)} } \\
        with\ 0<y<1,\mu=0.3,\sigma=4 
    \end{gathered}
\end{equation}

Using Equation~\ref{eq:bayes_orig} and Equation~\ref{eq:Ohki_2020}, the four probabilistic flood maps corresponding to different flood states were computed. To address unclassified pixels, the High-Resolution Land-Use and Land-Cover (HRLULC) map delineating rice paddy areas was subsequently utilized. In these HRLULC-identified rice paddy areas, unclassified pixels were reclassified as flooded if adjacent pixels, such as those representing ridges and roads, were flooded; otherwise, they were classified as non-flooded. Later, this study was further extended and reported in \cite{ohki2020automated}. The main concept of~\cite{ohki2020automated} is similar to that of~\cite{ohki2020study}, with the following two differences: 1) $F_0$ represents permanent no water pixels while $F_1$ represents permanent water pixels (non-flooded); 2) the prior probability was calculated using low-resolution hydrodynamic simulated data (FLDFRC):

\begin{equation}
    \begin{gathered}
        p(F_0)=p(F_1)=0.5(1-f) \\
        p(F_2)=p(F_3)=0.5f \\
        f=\frac{A}{1+{Be}^{{FLDFRC}_{max}}-C}  \\
        with A=0.5,\frac{1}{B}=0.005,C=0.2
    \end{gathered}
\end{equation}

The advantages of the methods proposed by~\cite{ohki2020study,ohki2020automated} are: 1) they account for unclassified pixels, which are often neglected by other flood mapping methods; and 2) they offer flexibility with input SAR features, allowing the number of classification categories to be adjusted based on the availability of SAR information.
 

Lastly, it is worth noting that the Bayesian method offers a straightforward solution for generating probabilistic flood maps. In addition to supplementing flood category maps, this is a crucial component for flood data assimilation, particularly hydrodynamic modeling. However, these methods have notable disadvantages: 1) the final results are heavily influenced by the prior distributions, which are based on various assumptions; 2) the computation can be quite expensive, as the distributions need to be parameterized when dealing with new input data or high-volume data.

\subsubsection{Machine learning classifiers}

Support Vector Machine (SVM), Quadratic Discriminant Analysis (QDA), K-Nearest Neighbours (KNN), and Stochastic Gradient Descent (SGD) are commonly used machine learning classifiers in SAR-based urban flood mapping \cite{aristizabal2020high, benoudjit2019novel, moya2020learning}. The study conducted by \cite{aristizabal2020high} utilised two distinct sets of characteristics. The first set consisted of intensity data acquired in both VV and VH, while the second set combined VV intensity, VH intensity, and HAND information. These feature sets were utilised to train three separate machine learning classifiers (i.e., SVM, QDA, and KNN), with each classifier employing specifically chosen sample points from three different study areas. Their findings demonstrated that SVM, QDA, and KNN exhibited modest variations in performance due to the utilisation of distinct decision boundaries. In addition, ~\cite{moya2020learning} employed SVM to create maps of urban floods. They utilised pre-event and co-event coherence as input characteristics. Both~\cite{aristizabal2020high, moya2020learning} have shown that they can effectively utilise well-trained machine learning models on various locations of the same event, as well as on new flood events, as long as the flood events in the training and test datasets share similar characteristics.

In addition, ~\cite{benoudjit2019novel} employed SGD in their study on flood mapping. The classifier was trained using water and land pixels obtained from pre-flood intensity data. It was then used to categorise post-flood intensity images into two classes, assuming that floodwater and permanent water bodies in backscatter have similar characteristics. The effectiveness of SGD in detecting urban floods was constrained by the insufficient optical imagery-derived training dataset, which did not generate accurate labels for the distinctive features of SAR data.~\cite{benoudjit2019novel}. The Random Forest (RF) algorithm has been employed for urban floodwater identification using multimodal SAR data, as well as for handling outliers in the training dataset~\cite{baghermanesh2022urban, chaabani2018flood}.~\cite{benoudjit2019novel} combined many data streams, including SAR data (i.e., intensity, coherence, phase, PolSAR, and SAR simulation), with other data such as topography and land cover maps. In their investigation,~\cite{chaabani2018flood} utilised RF to carry out flood detection. They divided the entire study region into ten distinct categories: river, dry agricultural, flooded agriculture, dry vegetation, flooded vegetation, dry urban zone, dry roads, dry grass, building roofs and flooded residential zone. The main distinguishing factors between flooded areas and other classes were intensity and coherence information. In addition to generating the final floodwater map, RF can also offer insights about the most effective inputs for performing the classification assignment. While RF produces classification results that are relatively accurate, its computation is resource-intensive due to the process of deriving final classifications by averaging the results from numerous decision trees. The latter study also highlights the potential of bistatic InSAR coherence from the TanDEM-X mission for more accurate flood mapping in complicated terrain, including urban and vegetated areas.

\subsubsection{Deep Learning}

The machine learning techniques previously discussed have produced numerous highly skilled models with transferable capabilities. However, these methods typically perform well on datasets of limited to moderate size. When classifying datasets with large volumes of data or high dimensionality, more advanced approaches, such as deep learning, should be considered. This is because deep learning techniques have the potential to effectively represent complex relationships within extensive datasets. To date, only three studies have explored the application of deep learning approaches for mapping urban floods using SAR data. 

\cite{zhao2022urban} employed the UNet architecture together with a novel urban-aware module that integrates urban-aware attention with channel attention. This module utilises a probabilistic urban mask derived from multi-temporal SAR data to enhance the network's comprehension of the most impactful features for categorising different types of floods, encompassing both bare and urban floods. The study utilised the probabilistic urban mask derived from SAR data, along with pre- and co-event coherence and pre- and post-event intensity data in both VV and VH polarisations. Although the proposed methods have achieved good accuracy in flood mapping, a notable limitation of the supervised deep learning approach is the time-consuming and challenging process of collecting training data for urban flood mapping, primarily due to the scarcity of available ground truth data. However, this study emphasizes large-scale flood mapping, where the model was trained on a small dataset but tested over broader areas, including regions with different flood events not present in the training data, as well as non-flooded areas. In contrast, other urban flood mapping techniques, such as Bayesian inference-based approaches, machine learning classifiers, and conventional methods, have not been evaluated across entire Sentinel-1 image frames (e.g., 250 km swath width) with various flood events across different countries or continents. Despite the limitation of the small training dataset, the Urban-aware UNet still demonstrates strong robustness.

In order to address the constraints presented by annotated data,~\cite{li2019urban} introduced a temporal-ensembling self-active learning convolutional neural network (A-SL CNN). The A-SL CNN was developed as a method for identifying urban floods by utilising SAR intensity and InSAR coherence. The study utilised a temporal-ensembling convolutional neural network (CNN) to produce pseudo-labels for the purpose of updating the training dataset. They iterated the procedure until they met a clearly defined stop condition. The A-SL CNN was trained using training samples from both real and pseudo-labels in order to generate accurate flood maps. This work demonstrated that the A-SL CNN model surpassed its supervised equivalent, suggesting that semi-supervised learning has great potential for urban flood mapping by overcoming the constraint of limited real annotations. Due to the potential for increased noise in the generated pseudo-labels for pixel-based segmentation, the authors decided to apply the semi-supervised model to image patches instead of individual pixels. Additional investigations are required to use semi-supervised learning for large-scale pixel-based urban flood mapping. 

Unsupervised learning was recently introduced to urban flood mapping by~\cite{yang2023promoting}. The researchers employed an unsupervised change detection technique that relied on generative adversarial networks (GAN) and out-of-distribution detection. This technique transforms SAR intensity data into three latent codes, which correspond to the surface conditions, incidence angle, and topography effects. The noise model is represented using a multimodal unsupervised image translation (MUNIT) model. The non-flooded synthesis SAR image was created using the SAR data provided, and flooded pixels were identified by detecting deviations from the typical distribution of non-flooded conditions. The key innovation of this approach lies in its ability to bypass the need for annotated flood pixels and instead focues on the information offered by a large number of non-flooded pixels. Flood pixels exhibiting variations in intensity, whether they are increasing or decreasing, can be quickly identified by comparing them with an actual SAR image alongside a non-flood SAR image that has been generated adversarially, as flooded pixels stand out as outliers compared to the non-flooded state. However, further investigation is required to evaluate the robustness and performance of this method across different flood events and various SAR sensors.

While the deep learning methods mentioned above have shown promise in urban flood mapping, several challenges remain regarding generalizability during the operational phase of large-scale flood mapping: 1) the Urban-aware UNet proposed in~\cite{zhao2022urban} heavily relies on the training dataset. Accurate flood mapping generation requires a well-trained model with a high-quality annotated dataset with significant diversity and volume, ensuring the model is able to identify the effective representation of the characteristics of various flooded and non-flooded conditions; 2) A similar challenge exists in the A-SL CNN model proposed by \cite{li2019urban}, where the quality and representativeness of labeled data are critical for accurate mapping. If the feature distribution of SAR data during inference differs from that of the labeled data, the method may fail; 3) the robustness of the GAN-based method proposed by~\cite{yang2023promoting} also needs further evaluation, as the radar noise pattern may be different in different images, with a possible explanation being that the information within the image patch may not adequately condition the noise model.

\subsection{Summary of different urban flood mapping methods} \label{sec:summary_method}

\textbf{Visual inspection} is a simple and straightforward method for identifying urban floods, despite its inherent limits in producing quantitative results. This approach can be useful for quickly assessing disaster situations in a qualitative manner, enabling rapid initiation of relief efforts after a catastrophe. However, to reduce the inherent ambiguity of SAR data and obtain precise, quantitative information on flood impacts, more sophisticated classification methods should be considered. 

\textbf{Fuzzy-logic based approaches} may easily incorporate various data streams from SAR data in conjunction with additional auxiliary data, such as digital elevation models (DEM) and land cover maps. They use expert-knowledge to determine essential initial values, while also allowing for precise adjustments of thresholds in specific circumstances. This approach can be utilised efficiently if the key variables influencing classification have been identified and the most optimal fuzzy rules are established. These approaches are suitable for addressing uncertainty related to the flood extent map, particularly when dealing with multiple sources of data.

\textbf{Region-growing-based approaches} necessitate substantial computational resources and heavily depend on the selection of initial seeds. They are most appropriate for flood scenarios that involve vast and continuous areas of flooding, rather than fragmented zones of inundation. High-resolution SAR data, such as 3 m TerraSAR-X and COSMO-SkyMed, are more suitable for capturing floodwater compared to medium-resolution SAR data, like 20 m Sentinel-1. The high-resolution data can capture more pixels, leading to the identification of continuous flooded areas that are more likely to exhibit spatial continuity in complex urban environments. In contrast, medium-resolution SAR images may confine floods to relatively small areas, making region-growing algorithms less effective for detecting urban floods.

\textbf{Decision tree-based approaches} are renowned for their simplicity, minimal computational requirements, interpretability, and capacity to learn from limited training data. However, they do not possess the ability for end-to-end trainability, as they make decisions at each individual node based on the data that is currently available, rather than optimising a global objective function. As a result, their ability to accurately identify intricate patterns in the data may be inferior to that of more advanced algorithms. These strategies are suitable for simple flood scenarios, when the causes contributing to the floods are well understood.

\textbf{Thresholding} techniques include both manual and automatic methods. The first option provides a straightforward and accurate approach to selecting the threshold, but requires experience and is subject to subjective judgments. In contrast, automatic method automates the threshold selection process, improving its reliability but also increasing the computional complexity. Thus, situations where the flood area is small may be more suited for manual thresholding, while automatic thresholding is likely to be more beneficial for large-scale flood mapping.
    
\textbf{Electromagnetic model}-based approaches have led to the development of a deterministic solution for urban flood mapping. The scattering model facilitates the direct inversion of various variables, such as water level. However, this approach entails certain limitations. The model's inherent complexity necessitates predetermination of a multitude of parameters, which introduces uncertainties from prior knowledge and model assumptions into the resulting urban flood map. When there is a requirement for flood water depth estimation, electromagnetic model-based methods may be considered. 

Researchers are exploring machine learning techniques to enhance the generalizability, robustness, and automation of SAR-based urban flood mapping methods, particularly in complex flood scenarios and with larger datasets.

\textbf{Bayesian inference-based approaches} are highly effective in incorporating prior knowledge and addressing uncertainty estimation. They offer more straightforward interpretations compared to neural networks. However, Bayesian approaches often necessitate stringent assumptions and encounter related difficulties, such as the curse of dimensionality. It is advisable to employ these approaches in situations where we possess substantial prior knowledge or when it is necessary to estimate the level of uncertainty associated with the flood map. Meanwhile, we can employ these techniques when we require a probabilistic urban flood map as opposed to a classification map.

\textbf{Machine learning classifiers} exhibit robust performance when applied to datasets with limited sample sizes, especially when combined with established feature engineering techniques. One of the major qualities of these classifiers is capacity to ensure excellent interpretability, allowing for a clear understanding of how features affect classification outcomes. However, their effectiveness diminishes when dealing with high-dimensional data and complex tasks. These classifiers usually require careful consideration of feature selection and the quality of representation. Hence, it is recommended to employ these techniques for managing small-scale datasets.

\textbf{Deep learning models} excel at processing large volumes of data for complex tasks. They require extensive training data and incur significant computational costs. Achieving optimal performance relies heavily on parameter tuning and optimisation, given the complex nature of the model structure. Therefore, once the model is well-trained with large and diverse datasets, we recommend deep learning-based methods for handling complex flood scenarios in the real world.

\section{Open Challenges, emerging technologies and future directions}
\label{sec:Challenges}

Although SAR-based urban flood mapping is currently receiving significant attention, there are still several unresolved issues in this field. This section provides a comprehensive examination of each aspect of the urban flood mapping process. It includes: (1) a detailed discussion of the SAR data involved and a brief overview of the current and future missions of SAR satellites, emphasizing the need for improved temporal/spatial resolution; (2) a critical examination of data pre-processing, such as terrain correction and speckle filtering, which are closely related to considerations of data quality; (3) an analysis of the challenges associated with current methodologies and their potential incorporation in future work; (4) an explanation of the difficulties in establishing a standardised benchmark dataset for urban flood mapping using SAR data; (5) a discussion of future directions, taking into account the TRLs and the feasibility of urban flood mapping techniques.

\subsection{Resolution of SAR data}

\begin{figure}[htb!]
     \centering
     \includegraphics[width=\linewidth]{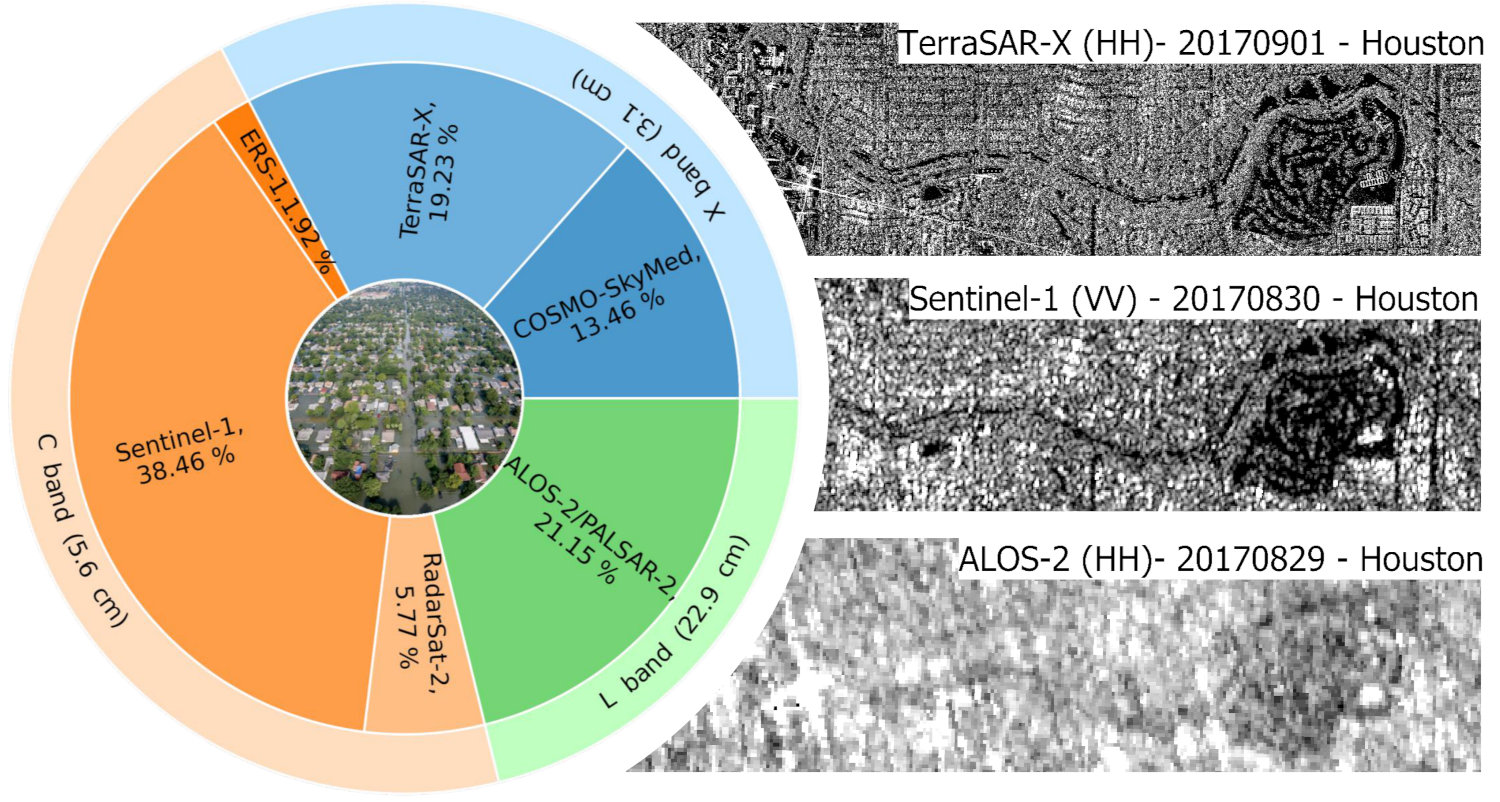}
\caption{SAR data acquired by different satellites in different bands considering the 49 selected studies.}
\label{fig:SAR_data_percentage}
\end{figure}

Table~\ref{tab:list_paper} shows that most current urban flood mapping studies use SAR data with medium to high spatial resolution. Approximately 40\% of the studies opted for high spatial resolution Sentinel-1 data (10-20 m) instead of very high spatial resolution TerraSAR-X, COSMO-SkyMed, and ALOS-2/PALSAR-2 data (1-8 m) due to the greater accessibility of Sentinel-1 data (Figure~\ref{fig:SAR_data_percentage}).

\begin{figure}[htb!]
     \centering
     \includegraphics[width=\linewidth]{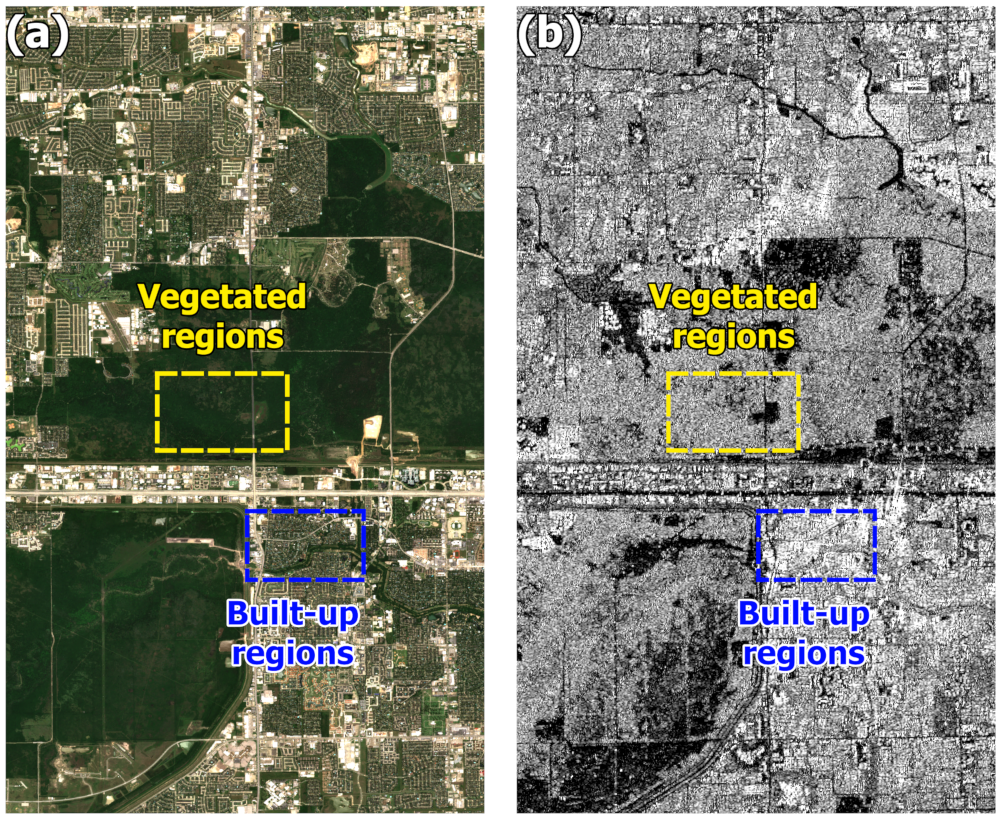}
\caption{The speckle noise in built-up regions and vegetated regions: (a) Sentinel-2 RGB composite image; (b) Sentinel-1 intensity image with speckle noise}.
\label{fig:noise}
\end{figure}

Nevertheless, it is imperative to acknowledge that spatial resolution plays an essential role in the process of using SAR technology for urban flood mapping. Although~\cite{lin2019urban} stated that accurate flood mapping in urban areas requires a ground resolution of at least 15 m, several studies still use Sentinel-1 intensity data with a resolution of 20 m. This phenomenon occurs as a result of the widespread use of speckle mitigation techniques such as multi-looking, which can cause a decrease in the spatial resolution of SAR data. Even in high-resolution SAR data, the process of multi-looking can cause the effects on the double bounce pattern to become blurred~\cite{iervolino2015flooding}. Speckle is an effect that occurs in SAR images when there are multiple scatterers randomly distributed within a resolution cell~\cite{moreira2013tutorial}. However, this impact is not significant in urban areas. Urban environments display a considerably higher number of bright pixels in comparison to places with bare soil and vegetation. The reason for this phenomenon is the double-bounce effect that occurs between the ground and building façades. In this effect, the speckle pattern does not fully develop due to the stable dihedral structures, as seen in Figure~\ref{fig:noise}. Within this framework, we argue that preserving Sentinel-1 data at its maximum spatial resolution, without considering speckle noise, has the potential to enhance the development of more accurate urban flood maps.

\begin{figure}[]
     \centering
     \includegraphics[width=\linewidth]{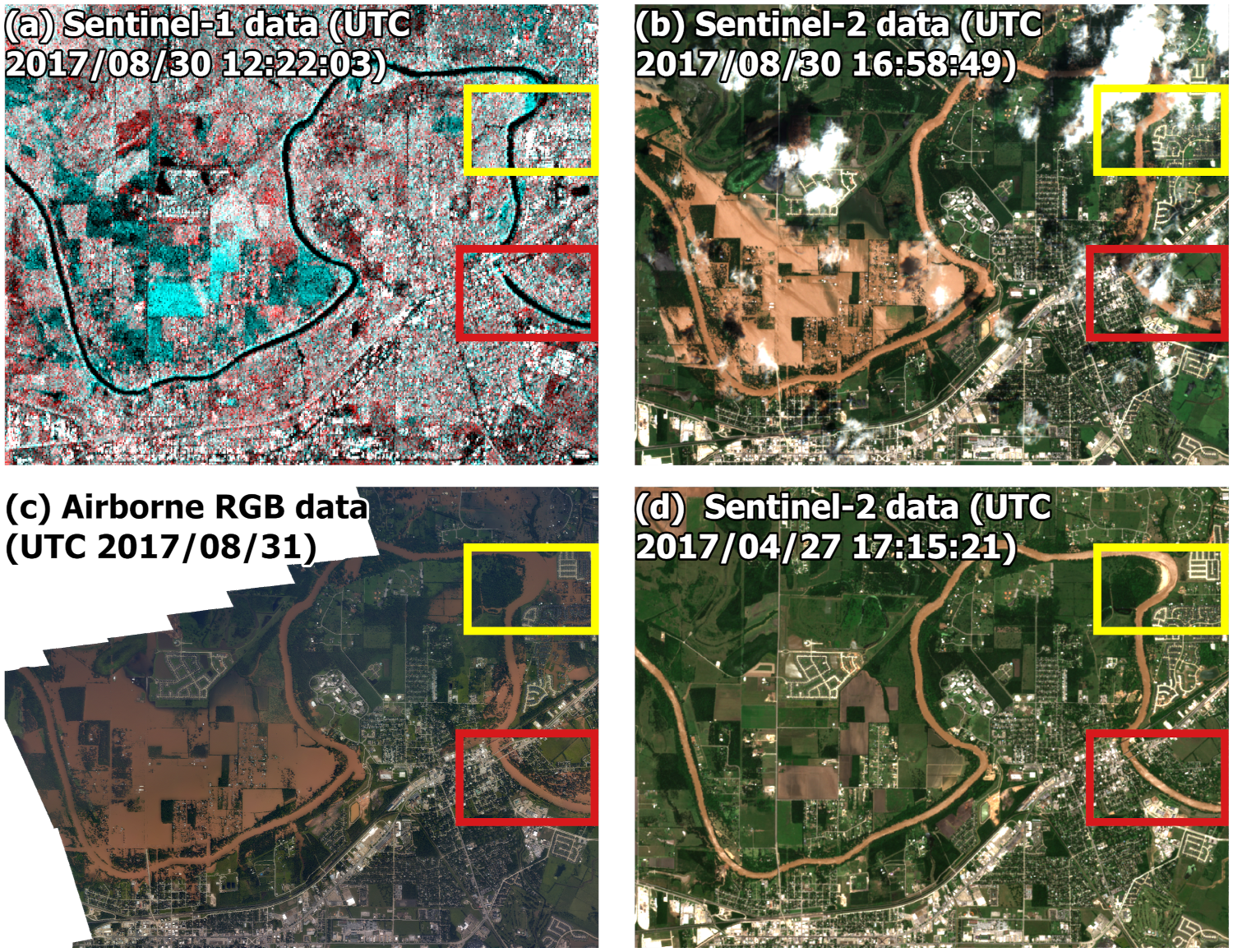}
\caption{Flood extent changes over time using the Houston flood as an example: (a) RGB composition from pre-event and post-event Sentinel-1 intensity data in VV polarization; (b) RGB composition from Sentinel-2 multispectral data obtained during flood event; (c) RGB image from NOAA’s airborne data; (d) RGB composition from Sentinel-2 multispectral data obtained in non-flood period}.
\label{fig:frequency}
\end{figure}

In addition, the exceptionally high spatial resolution of SAR data enables more in-depth analysis of floods in areas affected by shadow and layover. \cite{liu2018water} differentiate between the shadow of the building and floodwater by utilising a fuzzy-logic-based approach in the GF3 SAR intensity data. Simultaneously, \cite{mason2014detection} demonstrated the ability to identify urban floods in radar layover regions by comparing TerraSAR-X data with modelled double-bounce scattering mechanism. Other studies employed high-resolution DEMs and DSMs to differentiate betwen areas of shadow and flood water, thereby reducing the possibility of over- or underestimating the extent of flooding. However, it is not always possible to get centimetric-level DEMs or DSMs on a global scale. An effective approach to address this problem is utilising extremely detailed SAR data to enhance the resolution of current low-resolution DEMs or DSMs using deep learning techniques, as proposed by~\cite{tan2024rapid}. Moreover, the current SAR missions' temporal resolution poses a difficulty in mapping urban floods due to the rapid onset and receding nature of such floods. The impermeable surfaces in urban areas and the effectiveness of urban drainage systems further complicate this issue. In particular, the temporal resolution of SAR data is often insufficient for timely flood detection, typically exceeding one day (Figure~\ref{fig:SAR_data_percentage}). For example, during the Houston floods, Sentinel-1 captured the event, as illustrated in Figure~\ref{fig:frequency}, which shows the temporal changes in flood extent during late August 2017. Due of its temporal resolution of 3-4 days, Sentinel-1 usually captures only one image throughout an entire flood event in the Houston area. A comparison of Figure~\ref{fig:frequency} (b) and Figure~\ref{fig:frequency} (c) within the designated yellow box indicates that floodwater entered urban regions on August 31, 2017 (Figure~\ref{fig:frequency} (c)), leading to the lack of quantifiable flood data in the Sentinel-1 images (Figure~\ref{fig:frequency} (a)). In Figure~\ref{fig:frequency}, the red box shows a situation adjacent to the riverbank where Sentinel-1 data does not show any considerable floodwater, unlike the optical data from Sentinel-2 that was obtained just 4 hours later (Figure~\ref{fig:frequency} (b)).

\begin{figure}[]
     \centering
     \includegraphics[width=0.9\linewidth]{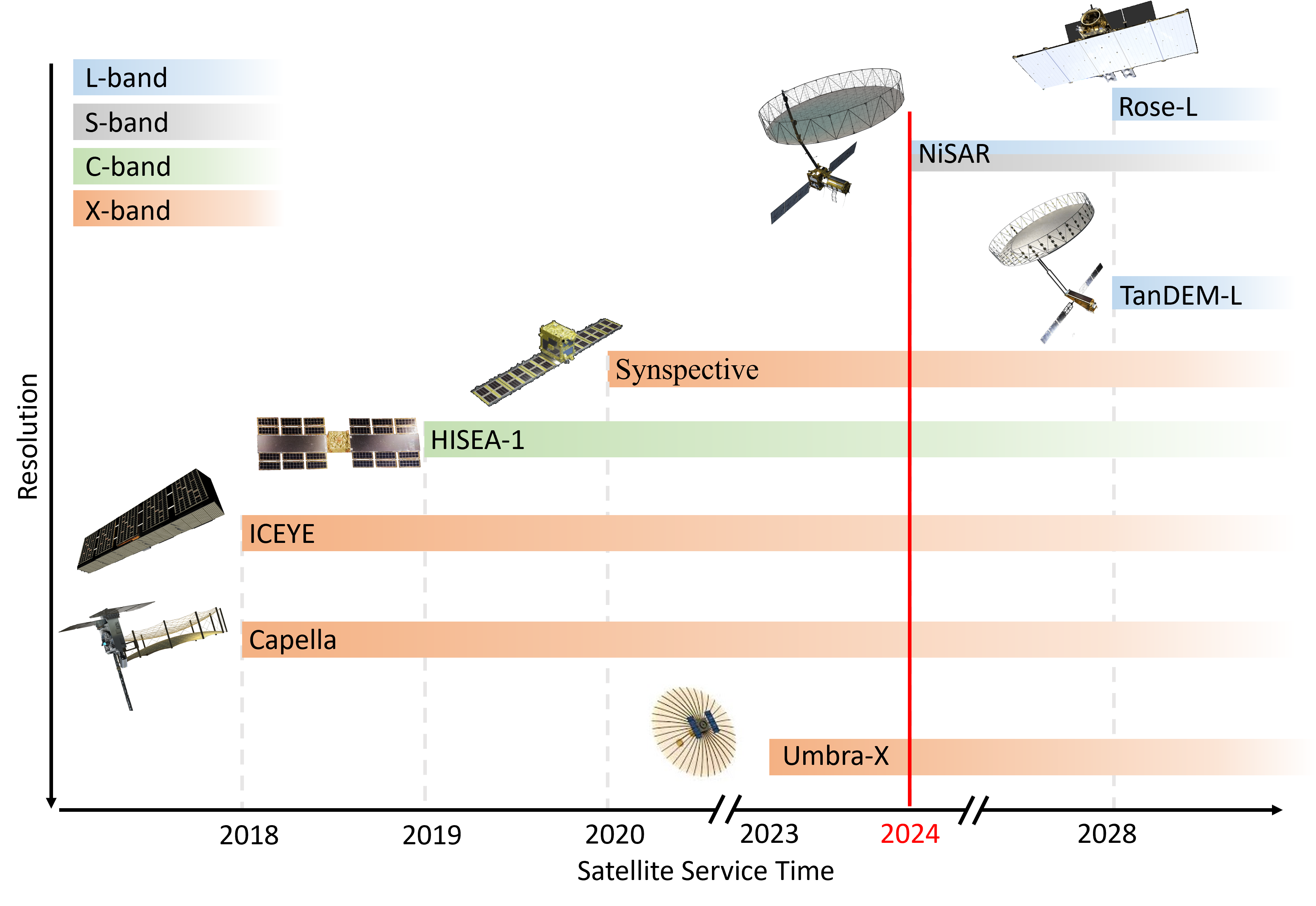}
\caption{NewSpace SAR satellites and future SAR satellites which could be used for urban flood mapping.}
\label{fig:newSat}
\end{figure}

NewSpace SAR satellites like ICEYE, HISEA-1, Capella, Synspective, Umbra-X, and other planned satellites have the potential to address the challenges posed by the low spatial and temporal resolution of current large SAR satellites (Figure~\ref{fig:newSat}). These advanced satellite technologies, characterised by their superior spatial and temporal resolution, offer promising avenues for future research in urban flood mapping. However, due to the satellite's payload limitations, ICEYE, HISEA-1, and Synspective can only capture data in VV polarization, while Capella and Umbra-X can provide SAR data in VV and HH polarizations. Consequently, relying exclusively on VV polarization may not adequately identify all flooded urban areas, as co-polarization may overlook important scattering mechanisms in urban flood mapping, such as double-bounce and multiple-bounce effects, which depend on the orientation of building façades relative to the SAR sensor’s LoS. These NewSpace satellites can provide meter-level spatial resolution, which helps describing complex urban scenarios better. Nevertheless, a thorough analysis of the trade-off between high spatial resolution and single-polarization acquisition is essential for effectively understanding how these factors influence the detection of urban floods. Another big challenge with these NewSpace SAR satellites is that the interferometric reference images are often not available because of their unstable flight orbit. Consequently, this data renders many current methods for mapping urban floods ineffective. It is noteworthy that ICEYE SAR data currently possesses a distinct advantage in offering a powerful urban flood indicator—InSAR coherence information. In contrast, other NewSpace satellites can only provide SAR intensity data. Furthermore, the European Space Agency (ESA), Thales Alenia Space, and the German Space Agency (DLR) are planning to launch the ROSE-L and TanDEM-L satellite in 2028, with the aim to diversify the dataset and accelerate research in this field. In addition to satellite SAR missions, UAV SAR could potentially address the issue of low temporal resolution~\cite{kundu2022flood}.

\subsection{Data pre-processing}

Data pre-processing constitutes a crucial step in influencing flood mapping results. Recognising that the side-looking nature of a SAR sensor inherently produces SAR images in the slant range, leading to rectangular pixels, is essential. Typically, the azimuth and slant range resolutions provide the SAR sensor's nominal spatial resolution. Pixels undergo resampling to squared pixels during the pre-processing phase, resulting in a reduction in spatial resolution. This kind of degradation is especially noticeable when using data with a resolution of one meter to ten meters in urban settings. This is because each pixel contains information from an area that is at least ten meters square (resolution-dependent), which includes different ground objects. \cite{sun2022large,recla2024sar2height} explored the application of slant-range SAR data for large-scale building height retrievals to maintain the full resolution of SAR data in remote sensing application. Utilizing slant-range geometry instead of ground-range geometry for SAR-based flood mapping could help preserve spatial resolution, particularly with high-resolution SAR data from sources like TerraSAR-X, COSMO-SkyMed, ALOS-2/PALSAR-2, and GF3.

In addition to reducing spatial resolution during terrain correction of SAR intensity data, the despeckling process is crucial for minimizing SAR noise, especially for subsequent applications such as flood mapping. SAR-related applications commonly employ despeckling filtering approaches, which, while not reducing the spatial resolution as a multilooking step, nonetheless influence the spatial resolution of SAR images. This, in turn, presents challenges for urban flood mapping using SAR data. The Lee filter \cite{lee1980digital} is the most popular intensity speckle filter in SAR-based flood mapping. It removes speckle noise by calculating a linear combination of the central pixel value and the average value of neighboring pixels within the moving kernel. Similarly, the Goldstein phase filter \cite{goldstein1998radar} serves as a standard phase filter in the context of InSAR coherence data. Both filters are widely used due to their good balance between feature preservation, despeckling capability, and computational efficiency, outperforming other available filters. To retain maximum information in SAR data with minimal resolution loss for flood mapping, advanced filters such as Block Matching and 3D Filtering (BM3D) \cite{dabov2007image}, Non-Local mean filtering \cite{deledalle2011nl,zhu2018potential}, and several deep learning-based filtering approaches \cite{franceschetti2002canonical} can be considered in the flood mapping domain. However, it is noteworthy that while speckle filters play an essential role in conventional flood mapping methods, some studies \cite{sun2022large,zhao2024IGARSS} report that speckle has a minimal impact on deep learning-based methods in SAR-related applications. It is believed that numerous convolution layers in convolutional neural networks (CNNs) contribute to noise removal \cite{sun2022large,zhao2024IGARSS}. Therefore, we recommend considering pre-processing procedures in conjunction with flood mapping approaches, as certain applications may not require SAR denoising.

\subsection{Methodology}

The expansion of SAR satellite constellations enables multiple satellites to capture a single flood event. However, most SAR-based flood mapping methods depend on change detection and are designed for single-source SAR data, specifically pre-event and post-event images obtained from the same SAR sensors. Consequently, these methods struggle to handle multi-sensor, multi-resolution, and multi-frequency SAR data. Currently, many studies are focused on change detection using multi-spatial resolution remote sensing images and are employing various strategies. For example, some approaches involve transforming multi-resolution image pairs into image pairs of the same resolution through resampling and co-registration, allowing deep learning models to learn the relationships between changed and unchanged features across both images \cite{zhang2016change}. \cite{wang2020deep} proposed a feature-level change detection method for multi-resolution optical images, which extracts the features of image pairs using a Siamese-based convolutional network, as an alternative to change detection using multi-resolution remote sensing imagery at the pixel level. Therefore, multi-source SAR data with different spatial resolutions holds great potential for mapping flooded urban areas, necessitating the development of generic change detection methods to overcome input data limitations and enhance the utilization of abundant information from various SAR satellites.

During the literature review for this study, it was surprising to find that nearly all research relied solely on spaceborne SAR data obtained in single or dual polarizations. One notable exception is a study by \cite{watanabe2009simultaneous}, which utilized fully polarimetric airborne SAR (i.e., PiSAR) data with a spatial resolution of 2.5 m to map urban floods. This study showcased the ability to identify flooded vegetated areas and urban regions using both intensity and entropy metrics. Furthermore, polarimetric decomposition methods, such as Yamaguchi four-component decomposition, Pauli decomposition, and Freeman-Durden decomposition—can analyze backscatter from various scattering mechanisms within fully polarimetric SAR data, particularly focusing on double-bounce scattering, which is a crucial characteristic of urban floods in SAR imagery. Consequently, exploring the use of polarimetric decompositions in urban flood mapping would be a valuable avenue for future research, especially considering the increasing number of fully polarimetric SAR satellites in orbit, even though operational satellites typically do not operate at full polarimetry (Table~\ref{tab:FullPolSAR}).

\begin{table}
\centering
\caption {\label{tab:FullPolSAR} Information for current fully polarimetric SAR satellites.\centering} 
\begin{center}
\begin{NiceTabular}{l|l|l}
\toprule[2pt] 
\hline
 Satellite & Country & Band\\
\hline

RADARSAT-2&	Canada&	C\\ \hline
TerraSAR-X/TanDEM-X	&Germany&	X\\ \hline
ALOS-2	&Japan&	L\\ \hline
SAOCOM	&Argentina&	L\\ \hline
GF3&	China&	C\\ \hline
L-SAR	&China	&L\\

\hline
\bottomrule[2pt] 
\end{NiceTabular}
\end{center}

\end{table}

Furthermore, it is important to note that most recent SAR-based urban flood mapping methods only produce classification maps that show flooded and non-flood classes, and further research is necessary to determine the reliability of these results. In this study's literature review, only five out of 49 studies were able to provide probabilistic flood maps instead of classification maps using Bayesian inference-based methods (section~\ref{sec:ML}). However, those studies only provided information about SAR data uncertainty, failing to account for model uncertainty as they classified flood conditions using backscatter distributions and specific assumptions. To address the unexplored uncertainty in urban flood mapping, various uncertainty estimation methods commonly used in deep learning could be beneficial. \cite{abdar2021review} presents a comprehensive review of uncertainty qualification (UQ) in deep learning, categorizing all UQ methods into two groups: Monte Carlo (MC) dropout \cite{gal2016dropout} and ensemble techniques. Currently, researchers widely introduce the MC dropout method in image segmentation and change detection using optical remote sensing imagery \cite{dechesne2021bayesian, kampffmeyer2016semantic,khoshboresh2021building}. Also, \cite{haas2021uncertainty} showed that deep ensembles (DE) are better at SAR-based road segmentation than MC dropout and deterministic models on both in-distribution and out-of-distribution data. Therefore, it is promising that we can consider these UQ methods in urban flood mapping, as they provide uncertainty from both data and models. Furthermore, incorporating complementary information is crucial for large-scale urban flood mapping in areas where SAR sensitivity is limited. Currently, two exclusion maps for Sentinel-1 data~\cite{zhao2021deriving, zhao2023preliminary, GloFAS} exist, tailored for large-scale flood mapping in open areas where only SAR intensity data is utilized. Moving forward, we can combine these products with SAR-based urban flood mapping techniques. At the same time, further exploration is needed for SAR-insensitive regions, considering factors like InSAR coherence and interferometric phase.

\subsection{Necessity for a Benchmark Dataset of urban flood mapping}

Acquisition of reliable groud-truth data is a fundamental requirement for urban flood mapping methods, irrespective of the specific approach used. This data is essential for evaluating results using conventional methods and generating annotation data for deep learning techniques. However, as mentioned in the previous section, urban floods are fast-moving and highly dynamic events, making it challenging to capture small urban flood events using satellites. Moreover, floods often coincide with severe weather conditions, such as heavy rainfall and strong winds, which severely limit the availability of cloud-free satellite optical imagery. As a result, there is a scarcity of ground-truth data for urban floods. Currently, local disaster management organisations like NOAA from the USA, GSI from Japan, ZKI from Germany, and the UN primarily obtain ground truth data, specifically optical data with centimetre-level spatial resolution, through airborne campaigns. The scarcity of ground truth data poses a significant challenge to urban flood mapping studies, particularly those utilising deep learning techniques. The difficulties in deriving a reliable benchmark dataset suggest that the scientific community should explore different approaches to derive them. A combination of hydrodynamic model results, land cover maps, expert visual inspection of multiple SARs and optical data, topographic data, and other secondary information could be a way to go.

While deep learning shows promise for SAR-based urban flood mapping, relatively few studies have explored semantic segmentation algorithms in this context. This limited investigation is primarily due to the challenge of acquiring a comprehensive and diverse training dataset. Although supervised learning has demonstrated high accuracy and robustness in some deep learning-based studies, its success in global-scale urban flood mapping is heavily dependent on the quality of the training data. In contrast, unsupervised learning presents a viable alternative, as it does not require a labelled dataset, as seen in the work of \cite{li2019urbanRS} and \cite{yang2023promoting}. However, the significant computational demands of this approach have limited its application in large-scale flood mapping, where rapid response is critical. Semi-supervised learning, as applied by \cite{li2019urban}, uses a small amount of labelled data combined with a larger set of unlabelled data. Yet, this method remains constrained by the limited diversity of its labelled dataset, leading to models that may lack the resilience and generalizability seen in fully supervised approaches. Moreover, the unique characteristics of urban flood events have led to the development of a variety of techniques, making it difficult to compare their relative strengths and weaknesses. Consequently, evaluating and comparing these methods is essential for understanding their effectiveness. This also underscores the urgent need for a standardized benchmark dataset to assess urban floods globally.

The current bottleneck in fostering the development of SAR-based urban flood mapping using deep learning is a lack of established benchmark datasets. Two benchmark dataset platforms, one from EarthNet (\url{https://earthnets.github.io/}) \cite{xiong2022earthnets} and the other from IEEE (\url{https://eod-grss-ieee.com/dataset-search}), provide access to several datasets related to floods. To the best of our knowledge, 11 flooding-related datasets are currently available. While six datasets primarily focus on open flood mapping with different data, including SAR, optical, and other auxiliary data (e.g., DEM and hydrography information), four datasets cover urban floods using optical data or a combination of optical and SAR data. Moreover, the recently published open-access S1GFloods dataset introduced by \cite{saleh2024dam} primarily focuses on flooded open areas, only partially covering urban floods. To the best of our knowledge, the newest UrbanSARFloods dataset developed by \cite{zhao2024urbansarfloods} was introduced covering 18 urban flood events globally using Sentinel-1 Single Look Complex (SLC) data. It is worth noting that, although several flood datasets include SAR data, they primarily consider SAR intensity and only one dataset (i.e., UrbanSARFloods) includes InSAR coherence. Thus, essential information from InSAR coherence and the interferometric phase have not been utilized. Furthermore, it is crucial to store the SAR-based urban flood mapping benchmark dataset in GeoTIFF format to maintain the unique characteristics of SAR data and its geolocation information. 

The current situation indicates a significant gap in benchmark datasets specially tailored for SAR-based urban flood mapping. Table~\ref{tab:flood_event_list} provides a summary of several well-known urban flood events along with corresponding Sentinel-1 scenes, aiming to fill this gap and offer valuable data insights to the urban flood mapping community. Future efforts will focus on releasing a processed dataset to facilitate further research and advance studies in this specialized area.

\begin{table*}[htb]
\centering
  \caption{Urban flood events acquired by Sentinel-1 with corresponding filenames from related studies}
  \label{tab:flood_event_list}
\small

\begin{NiceTabular}{c|c|c|c}
\toprule[2pt]

  \hline  
   \multirow{3}{*}{ID}     &	\multirow{3}{*}{Location}    &	Time  &	\multirow{3}{*}{Sentinel-1 SLC Scene}	 \\
      &	    &  (DD/MM	 &	 \\
     &	&	/YYYY) &	  \\ \hline

    1 &	Chennai (India)  &	05/03/2015	&  S1A{\_}IW{\_}SLC{\_}{\_}1SDV{\_}20150305T003124{\_}20150305T003151{\_}004889{\_}006184{\_}082F \\ \hline
    2 &	Sicily (Italy) &	02/11/2015 & S1A{\_}IW{\_}SLC{\_}{\_}1SDV{\_}20151102T050427{\_}20151102T050454{\_}008421{\_}00BE5F{\_}F34C \\ \hline
    3&	Houston (US)&	19/04/2016	& S1A{\_}IW{\_}SLC{\_}{\_}1SDV{\_}20160419T122235{\_}20160419T122305{\_}010890{\_}010501{\_}9776	\\ \hline
    4&	 Lumberton (US)    &	11/10/2016	 & S1A{\_}IW{\_}SLC{\_}{\_}1SDV{\_}20161011T231340{\_}20161011T231408{\_}013449{\_}0157D0{\_}909C\\ \hline	
    5&	 Kinston, US	& 12/10/2016 &	S1A{\_}IW{\_}SLC{\_}{\_}1SDV{\_}20161012T111514{\_}20161012T111544{\_}013456{\_}01580C{\_}B042 \\ \hline 
    \multirow{2}{*}{6}  &Alicante (Spain)	&13/03/2017&	S1A{\_}IW{\_}SLC{\_}{\_}1SDV{\_}20170313T175356{\_}20170313T175423{\_}015677{\_}019CB4{\_}0878 \\
        &   &  19/03/2017 &  S1B{\_}IW{\_}SLC{\_}{\_}1SDV{\_}20170319T175307{\_}20170319T175335{\_}004781{\_}008597{\_}2642    \\ \hline   
    7  & Houston (US)&	30/08/2017&	S1B{\_}IW{\_}SLC{\_}{\_}1SDV{\_}20170830T122203{\_}20170830T122233{\_}007169{\_}00CA2C{\_}C92C \\ \hline	
    8	& Beleweyne, Somalia	&08/05/2018	&    S1A{\_}IW{\_}SLC{\_}{\_}1SDV{\_}20180508T024535{\_}20180508T024602{\_}021807{\_}025A3C{\_}4F31	  \\ \hline
    9  &	Beira, Mozambique&	20/03/2019&	S1B{\_}IW{\_}SLC{\_}{\_}1SDV{\_}20190320T030812{\_}20190320T030839{\_}015432{\_}01CE6E{\_}E422	   \\ \hline
    10 &	Golestan, Iran &	29/03/2019    &	S1A{\_}IW{\_}SLC{\_}{\_}1SDV{\_}20190329T141951{\_}20190329T142019{\_}026554{\_}02F9EE{\_}EB4F   \\ \hline
    \multirow{3}{*}{11}   &	Gatineau/  &	\multirow{3}{*}{02/05/2019} &	\multirow{3}{*}{S1A{\_}IW{\_}SLC{\_}{\_}1SDV{\_}20190502T225218{\_}20190502T225245{\_}027055{\_}030C2D{\_}CE4A} \ \\
    &	Sainte-Marthe-sur-le-Lac,  &	  &	   \\
    &	Canada &	  &	\\ \hline
    12  &	Iwaki/Koriyama, Japan&	12/10/2019	&   S1B{\_}IW{\_}SLC{\_}{\_}1SDV{\_}20191012T204221{\_}20191012T204249{\_}018447{\_}022C0C{\_}9351   \\ \hline
    13  &	Ibaraki Prefurcture, Japan &	13/10/2019	&   S1A{\_}IW{\_}SLC{\_}{\_}1SDV{\_}20191013T083400{\_}20191013T083436{\_}029438{\_}035935{\_}9F3D  \\ \hline    
    14 &	Beleweyne, Somalia  &	30/10/2019	&  S1A{\_}IW{\_}SLC{\_}{\_}1SDV{\_}20191030T024549{\_}20191030T024616{\_}029682{\_}0361A2{\_}39BF	 \\ \hline
    15  & 	Fishlake, UK &	11/11/2019 &	S1B{\_}IW{\_}SLC{\_}{\_}1SDV{\_}20191111T174925{\_}20191111T174953{\_}018883{\_}0239D1{\_}4C01	 \\
    &   &   	14/11/2019	&   S1A{\_}IW{\_}SLC{\_}{\_}1SDV{\_}20191114T061423{\_}20191114T061450{\_}029903{\_}036957{\_}9164	 \\ \hline
    16  &	Jakarta, Indonesia &	02/01/2020	& S1A{\_}IW{\_}SLC{\_}{\_}1SDV{\_}20200102T111514{\_}20200102T111542{\_}030620{\_}038227{\_}C598 \\ \hline   
    17  &	Pontypridd, UK &	16/02/2020 &	S1A{\_}IW{\_}SLC{\_}{\_}1SDV{\_}20200216T063114{\_}20200216T063142{\_}031274{\_}0398EA{\_}EE84	 \\ \hline
    18    &  Beleweyne, Somalia  & 		21/05/2020	& S1A{\_}IW{\_}SLC{\_}{\_}1SDV{\_}20200521T024549{\_}20200521T024615{\_}032657{\_}03C847{\_}DA42	 \\ \hline
    19  &	Banjarmasin, Indonesia	&20/01/2021&	S1A{\_}IW{\_}SLC{\_}{\_}1SDV{\_}20210120T215959{\_}20210120T220026{\_}036227{\_}043FC4{\_}7CED  \\ \hline
    
\bottomrule[2pt]  
\end{NiceTabular} 
\end{table*}

\subsection{TRLs of current urban flood mapping techniques}

Research breakthroughs do not always translate into technology that is ready for immediate use in real-world applications and services. Algorithms usually pass through different development phases before becoming operational services~\cite{martinez2021futures}. To understand an algorithm's technological stage and its degree of maturity, the TRL of methods have been defined~\cite{mankins1995technology}. NASA developed the TRL measurement system in 1995 to evaluate a technology's maturity level. Initially applied to aeronautical and space projects, NASA has since expanded its scope to include all project types, from the conceptual stage to commercial deployment. The current TRL scale consists of nine levels, and each level represents the maturity of a technology's development, from a simple idea (level 1) to its full deployment on the market (level 9). Based on the discussion of all methodologies in Section~\ref{sec:method_section} and the summary of TRLs in Table~\ref{tab:TRLs},  an attempt was made to categorise and evaluate current satellite SAR-based urban flood mapping methods. According to their maturity, most of them are still in the prototype phase (TRL 4-5). This is largely due to the complexity of these models, which often involve numerous assumed parameters, such as those in electromagnetic models, and rely heavily on specific prior knowledge, including visual inspections, rule-based methods, and many machine learning-based approaches. To our knowledge, only two methods~\cite{chini2019sentinel,zhao2022urban} have advanced to a TRL of 6-7, having been implemented on cloud computing platforms and demonstrated across multiple real flood cases with tested generalizability in operational environments. These methods were deployed on the Wasdi platform (\url{https://www.wasdi.cloud/}) and evaluated through projects involving several end users.

One of the open challenges in this field is the transition of scientific research findings into practical industrial applications. To achieve this, it is necessary to not only validate the effectiveness and applicability of the developed technology in real-world scenarios, but also highlight the utility of urban flood mapping products on a global scale for society. The development of online platforms that offer near-real-time urban flood maps globally, akin to the Global Flood Monitoring (GloFAS GFM) (\url{https://global-flood.emergency.copernicus.eu/glofas-forecasting/}) and the Dartmouth Flood Observatory (\url{https://floodobservatory.colorado.edu/index.html}), is a promising direction for the future.

\begin{table*}[ht!]
\centering
\caption {\label{tab:TRLs} Summary of TRLs adapted from ~\cite{martinez2021futures}} 
\small
\begin{NiceTabular}{p{0.09\linewidth} | p{0.11\linewidth}| p{0.15\linewidth} | p{0.23\linewidth} | p{0.03\linewidth}| p{0.23\linewidth}}
\toprule[2pt]
\hline
Environment & Goal & Product /Evaluation & Outputs & TRL & Description \\ \hline
\multirow{3}{*}{Laboratory} & \multirow{3}{*}{Research} & \multirow{3}{*}{Proof of concept} & Scientific articles published on the principles of the new technology & 1 & Basic principles observed \\ \cline{4-6} 
 &  &  & Publications or references highlighting the applications of the new technology. & 2 & Technology concept formulated \\ \cline{4-6} 
 &  &  & Measurement of parameters in the laboratory & 3 & Experimental proof of concept with quantitative results \\ \hline
\multirow{3}{*}{Simulation} & \multirow{3}{*}{Development} & \multirow{4}{*}{Prototype} & Results of tests carried out in the laboratory & 4 & Technology validated using simulated flood cases \\ \cline{4-6} 
 &  &  & Components validated in a relevant environment. & 5 & Technology validated using relative small samples from real flood cases \\ \cline{4-6} 
 &  &  & Results of tests carried out at the prototype in a relevant environment. & 6 & Technology demonstrated in relative large numbers of real flood cases and tested generalizability \\ \cline{1-2} \cline{4-6} 
\multirow{3}{*}{Opertional} & \multirow{3}{*}{Implementation} &  & Result of the prototype level tests carried out in the operating environment. & 7 & System prototype demonstration in operational environment \\ \cline{3-6} 
 &  & Commercial (certified) product & Results of system tests in final configuration. & 8 & System complete and qualified \\ \cline{3-6} 
 &  & Deployed product & Final reports in working condition or actual mission. & 9 & Actual system proven in operational environment \\ \hline
\bottomrule[2pt]
\end{NiceTabular}
\end{table*}

\section{Social Implications}
\label{sec:Application}
Invaluable insights for a multiplicity of practical applications at both micro- and macro-perspective are provided by flood maps, particularly those that are custom-designed for urban areas. They are crucial for both private and public sector entities, including insurance companies, real estate developers, business enterprises, urban development departments, emergency management agencies, and environmental protection departments. Consumer purchasing decisions~\cite{reich2020informed}, developers' pricing strategies~\cite{bin2006real}, insurance companies' premium calculations and compensation processes~\cite{fema}, and various other aspects of decision-making in the realms of real estate and risk management are all influenced by flood risk information derived from these maps. 

As a public sector entity, the United Nations (UN) defined 17 Sustainable Development Goals (SDGs) in order to achieve peace and prosperity for all people by 2030, where satellite SAR data can play a crucial role in monitoring and achieving most of these goals~\cite{persello2022deep}. More specifically, flooding, as one of the most damaging hazards worldwide, can result in numerous negative consequences, including but not limited to damage to infrastructure, loss of lives and livelihoods, displacement of populations, contamination of water sources, and disruption of essential services. Moreover, there is an interdependent relationship between flooding and poverty: flooding exacerbates pre-existing poverty and amplifies the adverse effects of flooding by heightening vulnerability, while poverty itself could be a consequence of flooding~\cite{echendu2020impact}. For example, ~\cite{mtapuri2018flooding,brouwer2007socioeconomic} reported that the poor households impacted by the 2004 flooding in Bangladesh experienced a loss of more than twice their total income compared to affected non-poor households. Similarly, flooding destroyed millions of crops and killed millions of farm animals, fueling growing hunger~\cite{Somaiyah2023Pakistan}.~\cite{de2013soil} have investigated a reinforcing relationship between flooding and land degradation, demonstrating that flooding degrades land and land degradation increases the risk of flooding. Thus, flood mapping techniques are capable of providing information about the extent and depth of flood water, playing an important role in flood damage assessment and flood risk assessment. This, in turn, contributes to achieving SDGs such as No Poverty (SDG 1), Zero Hunger (SDG 2), and Life on Land (SDG 15). Recently, \cite{rentschler2023global} found that rapid urbanisation has generally increased in flood zones globally since 1985. Therefore, as a systemic risk, urban flooding may occur with higher frequency, resulting in more adverse consequences such as damaged infrastructure, environmental pollution, the spread of diseases, and social inequalities. Generally, urban floods can damage city infrastructure, including underground drainage systems, sewage pipes, sewage treatment plants, landfills, and chemical plants that produce toxic pollutants. The destruction of these facilities can lead to contamination of soil, surface water, and groundwater, resulting in environmental pollution~\cite{arrighi2018flood}. This pollution can further contribute to the spread of diseases~\cite{han2021urban,coalson2021complex,memon2014morbidity}. Consequently, urban flood maps can facilitate timely response measures in the event of urban flooding during the disaster relief phase. This includes evaluating post-flood health risks and identifying flooded polluted water bodies, supporting SDG goals such as Good Health and Well-being (SDG 3) and Clean Water and Sanitation (SDG 6).~\cite{rozer2022managing} suggests that precise information on floodwater extent and depth, available for damage assessments, can aid in the creation of flood-resilient and sustainable infrastructure, communities, and urban areas. This contributes to the achievement of Industry, Innovation, and Infrastructure (SDG 9) and Sustainable Cities and Communities (SDG 11). The global and regional flood exposure analysis reveals that 1.61 billion (89\%) of the world's flood-exposed people reside in low- and middle-income countries, while about 193 million (11\%) reside in high-income countries~\cite{rentschler2022flood}. Additionally, it shows that unequal countries tend to suffer more flood fatalities~\cite{lindersson2023wider}. Similar phenomena occur at the country or community level. For example, in the US, poorer communities are often disproportionately located in low-lying, flood-prone neighbourhoods with deficient infrastructure and limited access to shelter, as a direct result of decades of redlining and environmental racism~\cite{ermagun2024high}. Therefore, we can use urban flood maps as an efficient indicator for vulnerability assessments, thereby promoting the realisation of Reduced Inequalities (SDG 10). Climate change-induced heightened precipitation closely correlates with the escalation of flooding~\cite{Caretta2022water}. Urban flood mapping techniques serve as valuable tools for observing and measuring flooding, thereby enhancing understanding of the impacts of climate change and laying the groundwork for climate change mitigation efforts. This contributes to the achievement of Climate Action (SDG 13). Table~\ref{tab:SDG} summarises the selected SDGs, their targets, and indicators, providing an overview of how urban flood mapping contributes to monitoring and achieving the SDGs.

\begin{table*}[ht!]
\centering
\caption {\label{tab:SDG} The SDG targets and indicators that can be supported by urban flood map.\centering} 
\begin{NiceTabular}{l|l|l|l}
\toprule[2pt]
\hline
SDG & Targets & Indicators & Urban flood map in support of SDG targets and indicators \\ \hline
 \includegraphics[scale=0.2]{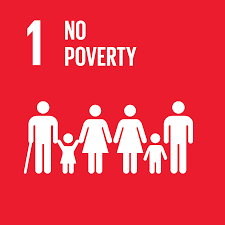} & 1.5 & 1.5.1 & Flood risk   assessment and post-flood damage assessment \\ \hline
 
 \includegraphics[scale=0.2]{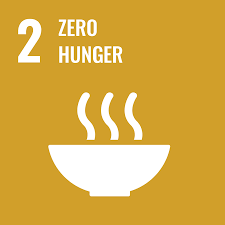} & \begin{tabular}[c]{@{}l@{}}2.3   \\ 2.4\end{tabular} & \begin{tabular}[c]{@{}l@{}}2.3.1   \\ 
 2.4.1\end{tabular} & Assessment of   vulnerability to flooding \\ \hline
 \includegraphics[scale=0.2]{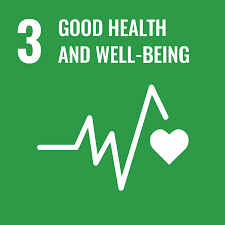} & \begin{tabular}[c]{@{}l@{}}3.1\\ 3.2\\ 3.3\\ 3.9\end{tabular} & \begin{tabular}[c]{@{}l@{}}3.1.1 \\ 3.3.3  \\ 3.9.3\end{tabular} & Assessment of   health risks caused by flooding \\ \hline
 \includegraphics[scale=0.2]{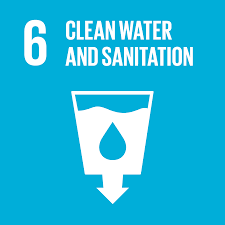} & \begin{tabular}[c]{@{}l@{}}6.5  \\ 6.6\end{tabular} & \begin{tabular}[c]{@{}l@{}}6.5.1  \\ 6.6.1\end{tabular} & Extraction of   flood polluted water bodies \\ \hline
 \includegraphics[scale=0.2]{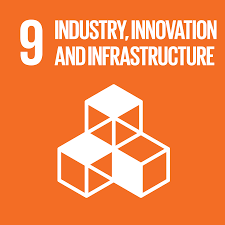} & 9.a & 9.a.1 & Assessment   of the flood resilience of infrastructure \\ \hline
 \includegraphics[scale=0.2]{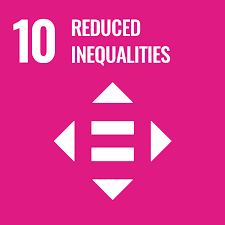} & \begin{tabular}[c]{@{}l@{}}10.3   \\ 10.4   \\ 10.b\end{tabular} & 10.4.2 & Assessment   of vulnerability caused by flooding \\ \hline
 \includegraphics[scale=0.2]{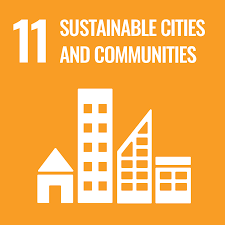} & \begin{tabular}[c]{@{}l@{}}11.5  \\ 11.b   \\ 11.c\end{tabular} & \begin{tabular}[c]{@{}l@{}}11.5.1  \\ 11.5.2\\ 11.5.3\end{tabular} & \begin{tabular}[c]{@{}l@{}}Flood risk assessment and early warning of vulnerable urban areas \\and flood-induced damage assessment. Information for the\\ development of resilient cities in developing countries.\end{tabular} \\ \hline
 \includegraphics[scale=0.2]{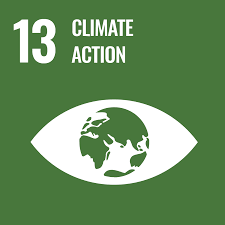} & \begin{tabular}[c]{@{}l@{}}13.1  \\ 13.2\end{tabular} &  & \begin{tabular}[c]{@{}l@{}}Flood risks and damage associated with climate change.\\ Measurement and observation of the Consequences of Climate Change\end{tabular} \\ \hline
 \includegraphics[scale=0.2]{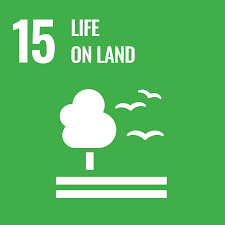} & 15.3 & 15.3.1 & Maps of flooded areas \\ \hline
\bottomrule[2pt]
\end{NiceTabular}
\end{table*}

Furthermore, evaluating vulnerability and conducting flood risk assessments are critical steps in flood risk management for individuals, the private sector, and the public sector. Individual customers may consider flood maps for their homes, workplaces, and other locations when purchasing flood insurance to protect their property~\cite{owusu2015public}. Furthermore, individuals can utilise flood hazard maps to define emergency plans, prepare resilient products such as waterproof doors, airbricks, and service duct covers, and identify reliable suppliers for property-level flood protection (PLFP)~\cite{owusu2015public}. Such information can also be considered in home buying or investment decisions~\cite{jung2018flood}. Moreover, flood maps play a crucial role in assisting insurance companies in determining compensation following a flood event. These maps provide detailed insights into the extent of flooding in specific areas, enabling insurers to accurately assess damages and calculate appropriate compensation for affected policyholders. By utilising flood maps, insurance companies can streamline claims processing, expedite settlements, and ensure equitable and efficient compensation for their clients impacted by floods~\cite{wu2023urban}. Flood maps offer a significant opportunity for public sectors such as urban development departments, emergency management agencies, and environmental protection departments to mitigate flood risks, enhance disaster preparedness, and bolster infrastructure resilience. Information on where floods could potentially occur, for example, facilitates urban flood resilience assessments, enabling planners and policymakers to analyse urban areas' ability to withstand and recover from flood events~\cite{bertilsson2019urban}. In addition to their importance in emergency response following a flood~\cite{mason2012near}, flood maps also serve as valuable tools for assessing and monitoring the impacts of floods on infrastructure integrity because floods can cause damage to the original structures and ground conditions, often resulting in hidden damage to buildings~\cite{prendergast2018structural}. Furthermore, we can use this information to educate the public, enhancing their understanding of flood risks and promoting community involvement in disaster preparedness initiatives.

\section{Conclusions}
\label{sec:conclusions}

In this study, we have undertaken a comprehensive overview of the current state-of-the-art SAR-based urban flood mapping. Our analysis covered five key areas: (1) The characteristics of urban floodwater as detected by SAR data; (2) The state-of-the-art methods for identifying urban floods, including flooded built-up regions and flooded residential areas using SAR data; (3) A review of benchmark datasets and the collection of urban flood events captured by Sentinel-1 in the 49 studies we examined; (4) Future challenges and research needs for urban flood mapping; and (5) Applications of SAR-based urban flood mapping. Our findings reveal that while InSAR coherence is a critical indicator for detecting urban floods, SAR intensity ($\sim$90\%) is the most widely used modality, whereas phase information ($\sim$8\%) remains underexplored. Furthermore, thresholding, particularly manual thresholding, is the most common method, while Bayesian inference is the leading machine learning based method in this field. 

Additionally, most studies produce flood classification maps ($\sim$90\%), with only a few ($\sim$10\%) incorporating uncertainty through Bayesian theory. Based on our review, we emphasize the need for further research into fully automated urban flood mapping, particularly in leveraging deep learning techniques, as many current methods still rely on empirical knowledge. Most studies focus on specific flood events, lacking robustness and generalizability. Moreover, the absence of a benchmark dataset that includes both SAR intensity and InSAR coherence and phase data has hindered thorough evaluation of SAR-based urban flood methods. Furthermore, our analysis of 49 previous studies reveals the predominant use of Sentinel-1 data, which is freely accessible through the Copernicus Programme. To support future research, we have compiled a collection of Sentinel-1 scenes depicting various urban flood events, facilitating easier data acquisition and intercomparison for researchers. We recommend that future studies utilise higher spatial resolution SAR data to gain a more comprehensive understanding of the spatial characteristics of floodwater in urban areas and to enhance our knowledge of the extent of affected areas. Besides, open access to satellite data for research purposes is critical, given the abundance of high-resolution SAR satellites and constellations in orbit. However, accessing such data for research purposes remains a challenge. In addition to high-spatial-resolution SAR data, future studies should also consider using change detection techniques utilising multi-source and multi-resolution SAR data to fully leverage the available information. Applying polarimetric decomposition techniques to fully polarimetric SAR data presents another promising research direction. Moreover, we need to further investigate the uncertainty of SAR-based flood maps to enhance our comprehension of their reliability. Finally, we must enhance the TRL of available urban flood mapping techniques to ensure their effectiveness in real-world situations. In conclusion, this study provides a detailed assessment of the current state of spaceborne SAR-based urban flood mapping, highlights the existing research gaps and challenges, and outlines the future directions needed to advance this field and make it more applicable in real-world flood management scenarios.

\vspace{11pt}

\bibliographystyle{IEEEtran}
\bibliography{reference.bib}

\vfill

\end{document}